\documentclass[twocolumn]{revtex4-1}
\usepackage{amssymb}
\usepackage{natbib}
\usepackage{graphicx}
\usepackage{amsmath}
\usepackage[bookmarks = false]{hyperref}
\usepackage{bm}
\usepackage[all]{hypcap}
\usepackage{graphicx}
\usepackage{colortbl}
\usepackage{booktabs}
\usepackage{enumerate}
\usepackage{dsfont}
\usepackage{enumitem}
\usepackage{braket}
\usepackage{mathrsfs}
\usepackage{multirow}
\usepackage{upgreek}
\usepackage{textcomp}
\newcommand{\degree}{^\circ}
\usepackage{soul}

\usepackage{array}
\newcommand{\PreserveBackslash}[1]{\let\temp=\\#1\let\\=\temp}
\newcolumntype{C}[1]{>{\PreserveBackslash\centering}p{#1}}
\newcolumntype{R}[1]{>{\PreserveBackslash\raggedleft}p{#1}}
\newcolumntype{L}[1]{>{\PreserveBackslash\raggedright}p{#1}}

\renewcommand{\andname}{\ignorespaces}

\newcommand{\msection}[1]{\vspace{\baselineskip}{\centering \textbf{#1}\\}\vspace{0.5\baselineskip}}

\begin{document}
\title{Entanglement of two quantum memories via fibers over dozens of kilometres}

\author{Yong Yu$^{1,\,2,\,*}$}
\author{Fei Ma$^{1,\,2,\,3,\,*}$}
\author{Xi-Yu Luo$^{1,\,2}$}
\author{Bo Jing$^{1,\,2}$}
\author{Peng-Fei Sun$^{1,\,2}$}
\author{Ren-Zhou Fang$^{1,\,2}$}
\author{Chao-Wei Yang$^{1,\,2}$}
\author{Hui Liu$^{1,\,2}$}
\author{Ming-Yang Zheng$^{3}$}
\author{Xiu-Ping Xie$^{3}$}
\author{Wei-Jun Zhang$^{4}$}
\author{Li-Xing You$^{4}$}
\author{Zhen Wang$^{4}$}
\author{Teng-Yun Chen$^{1,\,2}$}
\author{Qiang Zhang$^{1,\,2,\,3}$}
\author{Xiao-Hui Bao$^{1,\,2}$}
\author{Jian-Wei Pan$^{1,\,2}$}

\affiliation{$^1$Hefei National Laboratory for Physical Sciences at Microscale and Department
of Modern Physics, University of Science and Technology of China, Hefei,
Anhui 230026, China}
\affiliation{$^2$CAS Center for Excellence and Synergetic Innovation Center in Quantum
Information and Quantum Physics, University of Science and Technology of
China, Hefei, Anhui 230026, China}
\affiliation{$^3$Jinan Institute of Quantum Technology, Jinan, Shandong 250101, China}
\affiliation{$^4$State Key Laboratory of Functional Materials for Informatics, Shanghai Institute of Microsystem and Information Technology (SIMIT), Chinese Academy of Sciences, 865 Changning Road, Shanghai 200050, China}
\affiliation{$^*$These two authors contributed equally to this work.}

\maketitle

\textbf{Quantum internet\cite{Kimble2008,Wehner2018} will enable a number of revolutionary applications. It relies on entanglement of remote quantum memories over long distances. Despite enormous progresses\cite{Julsgaard2001,Chou2005,Moehring2007,Chou2007,Yuan2008,Hofmann2012,Bernien2013,Hensen2015,Delteil2015,Humphreys2018}  so far , the maximal physical separation achieved between two nodes is 1.3 km\cite{Hensen2015}, and challenges for long distance remain. Here we make a significant step forward by entangling two atomic ensembles in one lab via photon transmission through metropolitan-scale fibers. We use cavity enhancement to create bright atom-photon entanglement\cite{Simon2007,Bao2012,Yang2015a}, and harness quantum frequency conversion\cite{Kumar1990} to shift the atomic wavelength to telecom. We realize entanglement over 22~km field-deployed fibers via two-photon interference\cite{Simon2003,Zhao2007}, and entanglement over 50~km coiled fibers via single-photon interference\cite{Duan2001}. Our experiment can be extended to physically separated nodes with similar distance as a functional segment for atomic quantum networks, thus paving the way towards establishing atomic entanglement over many nodes and over much longer distance.}

Establishing remote entanglement is a central theme in quantum communication~\cite{Kimble2008,Yuan2010,Wehner2018}. So far, entangled photons have been distributed over long distance both in optical fiber~\cite{Inagaki2013} and in free space with the assistance of satellite~\cite{Yin2017}. In spite of these remarkable progresses, the distribution only succeeds with an extremely low probability due to severe transmission losses, and that photons have to be detected to verify the survival after transmission. Thus the distribution of entangled photons is not scalable to longer distance or to multiple nodes~\cite{Yuan2010,Sangouard2011}. A very promising solution is to prepare separate atom-photon entanglement in two remote nodes, and distribute the photons to a intermediate node for interference~\cite{Duan2001,Simon2003}. Proper measurement of the photons will project the atoms into a remote entangled state. Although the photons still undergo transmission losses, the success of remote atomic entanglement is heralded by the measurement of photons. Therefore, if the atomic states can be stored efficiently for a sufficient long duration, multiple pairs of heralded atomic entanglement can be further connected efficiently to extend entanglement to longer distance or over multiple quantum nodes through entanglement swapping~\cite{Sangouard2011}, thus making the quantum internet based applications feasible~\cite{Gottesman2012,Komar2014,Wehner2018}

Towards this goal, a great number of experimental investigations have been made with many different matter systems~\cite{Sangouard2011,Tittel2010,Duan2010,Reiserer2015,Aharonovich2016}, each of which has its own advantage in enabling different capabilities. To date, entanglement of two stationary qubits has been achieved with atomic ensembles~\cite{Julsgaard2001,Chou2005,Chou2007,Yuan2008}, single atoms~\cite{Hofmann2012}, NV centers~\cite{Bernien2013,Hensen2015,Humphreys2018}, quantum dots~\cite{Delteil2015}, trapped ions~\cite{Moehring2007}, etc. Nevertheless, for all systems, the maximum distance between two nodes which are physically separated remains 1.3 km~\cite{Hensen2015}. To extend the distance into metropolitan scale, there are three main experimental challenges, which are to achieve bright matter-photon entanglement, to reduce the transmission losses, and to realize stable and high-visibility interference in long fibers. In this paper we combine the atomic ensemble based quantum memory with efficient quantum frequency conversion (QFC)~\cite{Kumar1990}, and we realize the entanglement of two quantum memories via fiber transmission over dozens of kilometers. We make use of cavity enhancement to create a bright source of atom-photon entanglement. We employ the differential-frequency generation (DFG) process in a periodically-poled lithium niobate (PPLN) waveguide chip to shift the single-photon wavelength from near infrared to telecom O band for low-loss transmission in optical fibers. We then make use of a two-photon interference scheme~\cite{Simon2003,Zhao2007} to entangle two atomic ensembles over 22~km field-deployed fibers. Moreover, we make use of a single-photon interference scheme~\cite{Duan2001} to entangle two atomic ensembles over 50~km coiled fibers. Our work can be extended to long-distance separated nodes as a functional segment for atomic quantum networks and quantum repeaters~\cite{Briegel1998}, and is soon to enable repeater-based quantum communications, and paves the way towards building large-scale quantum networks over long distance in a scalable way in the near future~\cite{Kimble2008,Wehner2018}.

\begin{figure*}[hbtp]
	\centering
	\includegraphics[width=\textwidth]{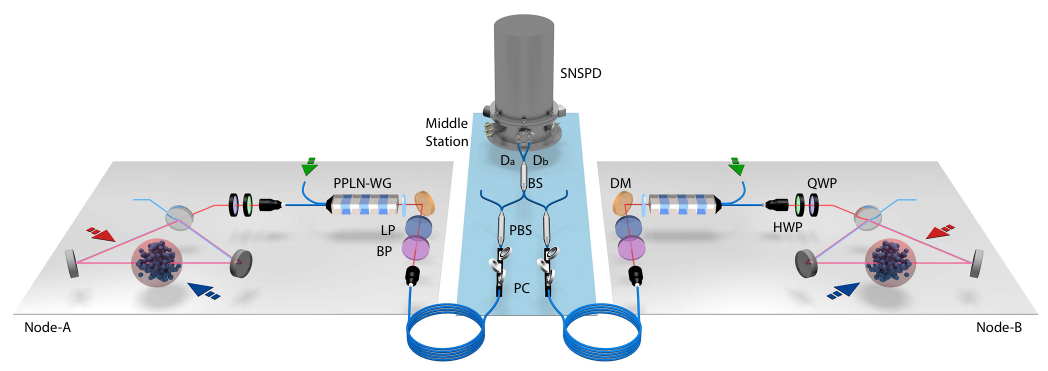}
     \caption{\textbf{Schematic of the remote entanglement generation between atomic ensembles.} Two quantum memory nodes (Node-A and Node-B in one lab) are linked by fibers to a middle station for photon measurement. In each node, a $^{87}$Rb atomic ensemble is put inside a ring cavity. All atoms are prepared in the ground state at first. We first create a local entanglement between atomic ensemble and a write photon by applying a write pulse (blue arrow). Then the write-out photon is collected along clockwise (anticlockwise) cavity mode and by sent to the QFC module. With the help of a PPLN waveguide chip (PPLN-WG) and a 1950~nm pump laser (green arrow), 795~nm write-out photon is converted to telecom O band. The combination of a half-wave-plate (HWP) and a quarter-wave-plate (QWP) helps coupling with the TM-polarized mode of the waveguide. After noise filtering, two write-out photons are transmitted through long fibers, interfered in a BS and detected by two SNSPDs with efficiencies of about $50\%$ at a dark count rate of 100~Hz. The effective interference in the middle station heralds two ensembles entangled. Fiber polarization controllers (PCs) and polarization beamsplitters (PBSs) before the interference BS is designed for actively compensating polarization drifts in the long fiber. To retrieve the atom state, we apply a read pulse (red arrow) counter-propagating to the write pulse. With the help of phase match of spin-wave and cavity enhancement, the atomic state is retrieved efficiently into anticlockwise (clockwise) mode of ring cavity.}
	\label{fig:setup}
\end{figure*}

\msection{Quantum memory with telecom interface}\label{sec:node}

Our experiment consists of two similar nodes linked via long-distance fibers, as shown in Fig.~\ref{fig:setup}. In each node, an ensemble of $\sim10^8$ atoms trapped and cooled by laser beams serves as the quantum memory~\cite{Sangouard2011}. All atoms are initially prepared in the ground state $\ket{g}$. Following the Duan-Lukin-Cirac-Zoller (DLCZ) protocol~\cite{Duan2001}, in each trial, a weak write pulse coupling  ground state atoms to the excited state $\ket{e}$ induces a spontaneous Raman scattered write-out photon together with a collective excitation of the atomic ensemble in a stable state $\ket{s}$ with a small probability $\chi$. The collective excitation can be stored for long duration and later be retrieved on demand as a read-out photon in a phase-matching mode by applying the read pulse which couples the transition of $\ket{s}\leftrightarrow\ket{e}$. The write-out and the read-out photons are nonclassically correlated. By employing a second Raman scattering channel $\ket{g}\rightarrow\ket{e}\rightarrow\ket{s'}$, we can create the entanglement between the polarization of the write-out photon and internal state ($\ket{s}$ or $\ket{s'}$) of the atomic ensemble~\cite{Yang2015a,Jing2019}. To further enhance the readout efficiency~\cite{Simon2007} and suppress noise from control beams, we build a ring cavity with a finesse of $F=23.5$ around the atomic ensemble. The ring cavity not only enhances the retrieval but also serves as a filter to eliminate the necessity of using external frequency filters to suppress noise~\cite{Jing2019}.

\begin{figure}[hbtp]
	\centering
	\
	\includegraphics[width=.8\columnwidth]{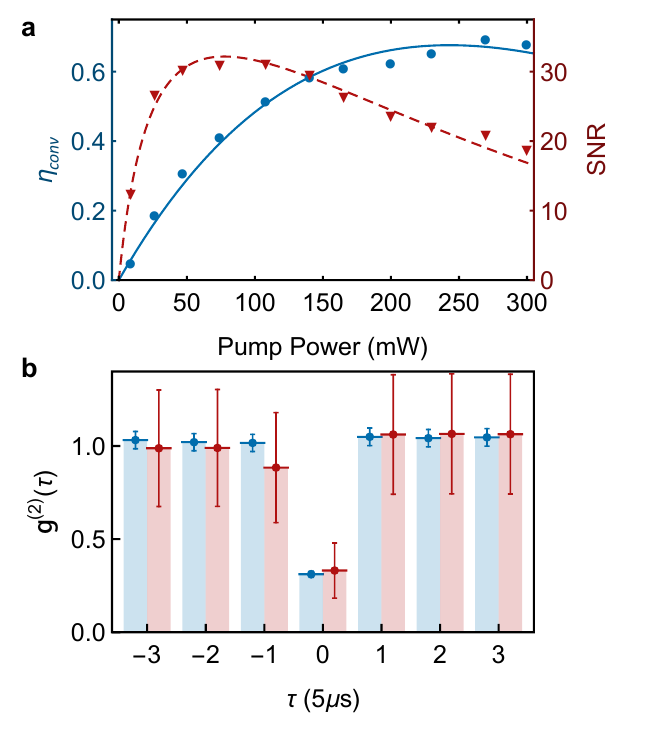}
	\caption{\textbf{Performance of the telecom interface.} \textbf{a}, The conversion efficiency $\eta_{conv}$ and SNR vary as a function of pump laser power. Blue dots refer to the overall conversion efficiency of the PPLN waveguide chip, and red triangles refer to SNR at $\chi=0.015$. \textbf{b}, Results of the Hanbury-Brown-Twiss experiment with (red) and without (blue) QFC at $\chi=0.057$. The write-out photons are measured conditionally on the detection of a corresponding read-out photon.}
	\label{fig:telecom}
\end{figure}

\begin{figure}[hbtp]
	\centering
	\
	\includegraphics[width=\columnwidth]{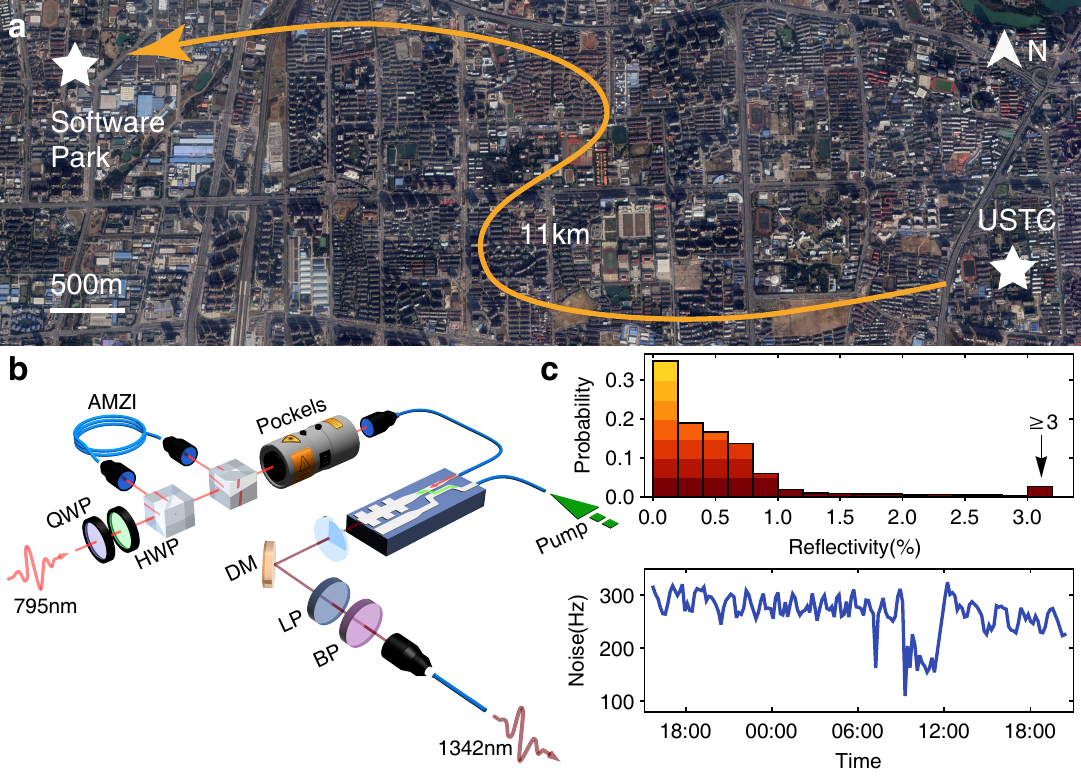}
	\caption{\textbf{Entanglement over field fibers.} \textbf{a}, Bird's-eye view of remote entanglement experiment over the field fiber. Two quantum nodes are located in USTC. Telecom photons from two nodes are transmitted through two parallel field-deployed fibers to the middle station located at Software Park of Hefei. Each fiber is 11~km long and has an 4~dB attenuation for 1342~nm photon. Map data: Google, Maxar Technologies. \textbf{b}, Setup for polarization photon QFC. Two PBSs and a coiled polarization maintaining (PM) delay fiber constitute an AMZI. Two orthogonal polarization components ($\ket{\circlearrowleft}/\ket{\circlearrowright}$) of 795~nm photon are separated in time domain after the AMZI, and the polarization information is actively erased by a Pockels cell. Then the time-bin encoded photon is sent to the QFC module. \textbf{c}, Probability distribution of the reflectivity for the polarization filtering PBS (shown in Fig.~\ref{fig:setup} after long fibers), with active compensation. The data shown was recorded once per second and accumulated during 24~hours. \textbf{d}, Background noise in SNSPD during 24~hours. Map data: Google Earth.}
	\label{fig:map}
\end{figure}

To create remote atomic entanglement over a long distance, it is crucial that the photons are suitable for low-loss transmission in optical fibers. Thus we shift wavelength of the write-out photon from near infrared (3.5~dB/km at 795~nm) to telecom O band (0.3~dB/km at 1342~nm) via the DFG process. We make use of reverse-proton-exchange PPLN waveguide chips. Optimal coupling efficiency and transmission for 795~nm signal and 1950~nm pump are simultaneously achieved in one chip by an integrated structure consisting two waveguides (see Fig.~\ref{fig:map}b and Supplementary Information). Fig.~\ref{fig:telecom}a shows that its conversion efficiency is up to $\eta_{conv}\approx70\%$ with 270~mW pump laser. During the conversion, there are three main spectral components of noise: 1950~nm, 975~nm and 650~nm, which come from pump laser and its second and third harmonic generation. They are all spectrally far enough away from 1342~nm and thus can be cut off via the combination of two dichroic mirrors and a long-pass filter edged at 1150~nm. The pump laser also induces broadband Raman noise, and the spectral brightness of which around 1342~nm is measured to be $\sim$500~Hz/nm. Thus, we use a bandpass filter (centred at 1342~nm, 5~nm linewidth) to confine this noise to $\sim$2.5~kHz, which corresponds to a signal-to-noise ratio of SNR$>20:1$ as depicted in Fig.~\ref{fig:telecom}a. The filtering process induces only $20\%$ loss, and fiber coupling brings extra $40\%$ loss. The end-to-end efficiency of our QFC module is $\eta_{QFC}=33\%$, which is the highest value for all memory-telecom quantum interfaces~\cite{Radnaev2010,DeGreve2012b,Maring2017,Bock2017,Ikuta2017,Walker2017,Dreau2018} reported so far to the best of our knowledge. In addition, we perform a Hanbury-Brown-Twiss experiment for the write-out photons with and without QFC, with the results shown in Fig.~\ref{fig:telecom}b, which verify that the single-photon quality is well preserved during QFC.

\begin{figure}[hbtp]
	\centering
	\includegraphics[width=\columnwidth]{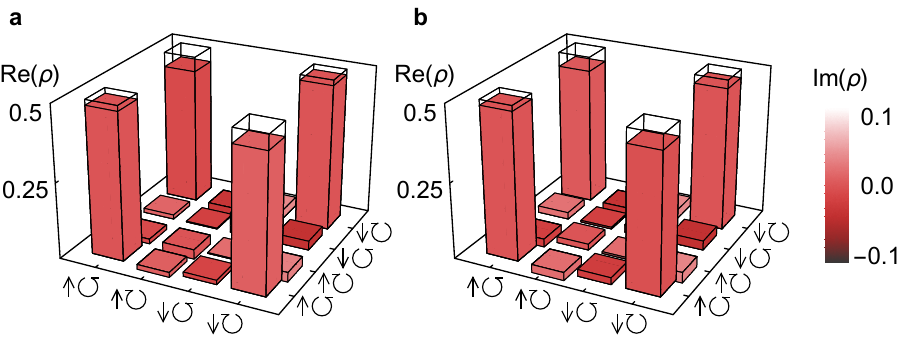}
	\caption{\textbf{Tomography of the atom-photon entanglement.} \textbf{a}, \textbf{b}, The reconstructed density matrix between write-out photon and atomic spin-wave in Node-A (\textbf{a}) and B (\textbf{b}). In each element of the matrix, the height of the bar represents its real part and the color represents its imaginary part. The transparent bars indicate the ideal density matrix of the maximally entangled state.
	}
	\label{fig:single}
\end{figure}

\msection{Entanglement over 22~km field fibers}
We first perform a two-node experiment via two-photon interference (TPI)~\cite{Zhao2007}. In each node, we create entanglement between polarization of the write-out photon and internal state of the collective excitation via a double-$\Lambda$ scheme (see Supplementary Information for level details). The entangled state can be expressed as $(\ket{\uparrow\circlearrowleft}+\ket{\downarrow\circlearrowright})/\sqrt{2}$, where $\ket{\uparrow}$ or $\ket{\downarrow}$ denotes an atomic excitation in $\ket{s}$ or $\ket{s'}$ respectively, and $\ket{\circlearrowleft}$ and $\ket{\circlearrowright}$ denote polarization of the write-out photon. To characterize the atom-photon entanglement, we perform quantum state tomography, with the result shown in Fig.~\ref{fig:single}. We get a fidelity of 0.930(6) for node A and $0.933(6)$ for node B when $\chi=0.019$. The two nodes are located in one lab in USTC east campus (N~$31^{\circ}50^{'}6.96^{''}$, E~$117^{\circ}15^{'}52.07^{''}$) as shown in Fig.~\ref{fig:map}a. Once the polarization entanglement is ready, the write-out photon is converted by QFC into telecom band locally. Two photons from different nodes are transmitted along two parallel field-deployed commercial fiber channels (11km/channel) from USTC to Hefei Software Park (N~$31^{\circ}51^{'}6.01^{''}$, E~$117^{\circ}11^{'}54.72^{''}$) as shown in Fig.~\ref{fig:map}a. Over there, we perform a Bell-state measurement (BSM) by detecting two photons simultaneously with superconducting nanowire single photon detectors (SNSPDs). A successful BSM result heraldedly projects the two atomic ensembles into a maximally entangled state
\begin{equation}
|\varPsi^{\pm}\rangle_{tpi}=\frac{1}{\sqrt{2}}(\ket{\uparrow}_A\ket{\downarrow}_B\pm\ket{\downarrow}_A\ket{\uparrow}_B),
\end{equation}
with a internal sign determined by the measurement outcome of BSM.

Strong polarization dependence of DFG in PPLN waveguide makes it difficult to perform QFC directly for a polarization encoded photon. In this experiment, we transform the polarization encoding into time-bin encoding~\cite{Farrera2018} and let the two photonic modes pass through the QFC module in sequence with the same polarization. As shown in Fig.~\ref{fig:map}b, the transformation is realized through an asymmetric Mach-Zender interferometer (AMZI) and a fast Pockels cell which erases the polarization distinguishability. For the time-bin encoding, it is crucial that the two modes have a stable relative phase shift, which is realized via active stabilization of the two AMZIs. Moreover, the transformation into time-bin encoding offers additional advantage of robustness in long-distance transmission in fibers.

\begin{figure}[hbtp]
	\centering
	\
	\includegraphics[width=0.7\columnwidth]{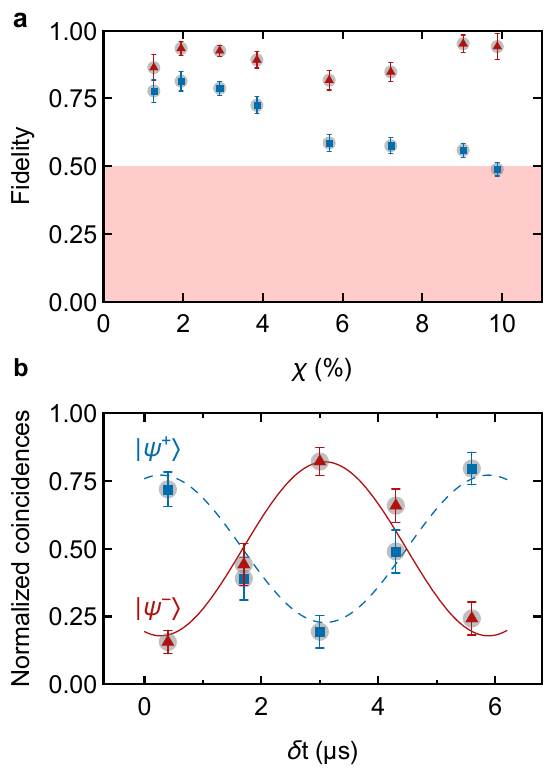}
	\caption{\textbf{Characterization of the remote entanglement via TPI.} \textbf{a}, Average fidelity of the remote entanglement $\ket{\varPsi^{\pm}}_{tpi}$ generated locally as a function of $\chi$. Blue (square) dots refer to the measurement result. Red (triangle) dots show the corrected results through deduction of accidental coincidences (see Supplementary Information). \textbf{b}, Normalized coincidences measured in the $\ket{\pm}=\ket{\uparrow}\pm\ket{\downarrow}$ basis for the two atomic qubits. The Raman pulse in Node-A is applied slightly later than Node B with an offset of $\delta t$, which induces a linearly changing phase in $\ket{\varPsi^{\pm}}$ and results in the observed oscillations. Parallel correlations ($\ket{+}\ket{+}$ or $\ket{-}\ket{-}$) of $\ket{\varPsi^+}$ (blue squares) and $\ket{\varPsi^-}$ (red triangles) are shown. Solid (red) and dashed (blue) lines correspond to the fitting results. The 5.4~$\upmu$s oscillation period agrees with Zeeman splitting between $\ket{\uparrow}$ and $\ket{\downarrow}$. This plot is based on $2.9\times10^4$ heralding events during a total measurement time of 487 hours over a period of 30 days. The error bars represent one standard deviation.}
	\label{fig:bdcz}
\end{figure}

Before long-fiber experiments, we characterize the atom-atom entanglement locally without QFC. For the measurement of the atomic qubits, we first apply Raman rotations~\cite{Jiang2016}, then we retrieve the excitations into read-out photons and make polarization measurement~\cite{Jing2019}. Measurement in arbitrary basis is realized via configuring the Raman pulses. Fig.~\ref{fig:bdcz}a shows the measured fidelity averaged for $\ket{\varPsi^\pm}$ as a function of $\chi$. At $\chi\simeq2\%$, we get $\mathcal{F}=0.798\pm0.063$ for $\ket{\varPsi^+}$ and $0.829\pm0.036$ for $\ket{\varPsi^-}$ respectively, which are in good agreement with the theoretical estimation. Furthermore, the fidelity is basically invariant along with $\chi$ after subtracting the accidental coincidences that is mainly due to high-order excitations in the Raman scattering process.

The field-deployed long fiber (L=22~km) induces 8~dB of attenuation. Besides, the long fiber leads to random rotations of polarization. To optimize the indistinguishability, we apply polarization filtering for the photons after long fiber transmission before BSM. In addition, to get a high filtering efficiency, we perform active polarization compensation by replacing the manual PCs in Fig.~\ref{fig:setup} with electric polarization controllers (EPCs) and minimizing reflections of the filtering PBSs. We get an average efficiency of $98\%$ as shown in Fig.~\ref{fig:map}c. To reduce the background noise in the fiber channels, we carefully shade all the fusion points and get an average background noise of $\sim$280~Hz (including dark counts of the detector). In the long fiber case, to increase the count rate, we set the excitation probability to $\chi=0.038$ and perform entanglement verification in a delayed-choice fashion~\cite{Ma2012}. The measured visibility in the $\ket{\uparrow}/\ket{\downarrow}$ basis is $V_1=0.684\pm0.075$ for $\ket{\varPsi^{+}}$ and $V_1=0.635\pm0.075$ for $\ket{\varPsi^{-}}$. Adjusting the Raman pulse delay $\delta t$, we could observe a sinusoidal oscillation in the $\ket{\uparrow}\pm\ket{\downarrow}$ basis as shown in Fig.~\ref{fig:bdcz}b with a visibility of $V_2=0.574\pm0.064$ for $\ket{\varPsi^{+}}$ and $V_2=0.647\pm0.066$ for $\ket{\varPsi^{-}}$. By assuming a similar visibility in the $\ket{\uparrow}\pm i\ket{\downarrow}$ basis, the entanglement fidelity can be estimated as~\cite{Guhne2009} $\mathcal{F}\simeq\frac{1}{4}(1+V_1+2V_2)=0.708\pm0.037$ for $\ket{\varPsi^{+}}$, and $0.732\pm0.038$ for $\ket{\varPsi^{-}}$, which significantly exceed the bound of $\mathcal{F}>0.5$ to witness entanglement for a Bell state. The measured heralding rate is $P_{her}=1.46\times10^{-6}$, half of which is due to double-excitation events from a single node. Thus the entangling probability is estimated to be $P_{ent} \simeq P_{her}/2 =0.73\times10^{-6}$.

\msection{Entanglement over 50~km coiled fibers}\label{sec:DLCZ}
The entangling probability in the TPI experiment~\cite{Zhao2007} is low since it scales as squared of $\chi$ and $\eta_{L/2}$, with $\eta_{L/2}$ being the overall optical efficiency from one node to the BSM. In contrast, a single-photon interference (SPI) scheme~\cite{Duan2001} gives an entangling probability which scales linearly as a function of $\chi$ and $\eta_{L/2}$. Thus, targeting a much higher entangling probability, we perform another two-node experiment via SPI. As shown in Fig.~\ref{fig:setup}, two pairs of Fock-state entanglement are created at Node-A and B respectively in the form of $\ket{0}_p\ket{0}_a+\sqrt{\chi}\ket{1}_p\ket{1}_a$, where 0 and 1 represent the number of photons or atomic excitations. Then the frequency converted photons from both nodes are transmitted along a long fiber, later combined through a fiber beamsplitter (BS) to perform SPI and eliminate its ``which way" information, finally detected with SNSPDs. A click from $D_a$ or $D_b$ heralds that two ensembles are mapped into a maximally entangled state
\begin{equation}\label{eq:psi}
|\varPsi^{\pm}\rangle_{spi}=\frac{1}{\sqrt{2}}(\ket{0}_A\ket{1}_B\pm e^{i\Delta\phi_{wo}}\ket{1}_A\ket{0}_B),
\end{equation}
where $\Delta\phi_{wo}$ is the accumulated phase difference between two fiber channels. To keep $\Delta\phi_{wo}$ in Eq.~\ref{eq:psi} constant, we harness an intermittent phase-locking loop in situ during every experimental interval to eliminate phase drift (see Supplementary Information).

\begin{figure}[hbtp]
	\centering
	\
	\includegraphics[width=0.7\columnwidth]{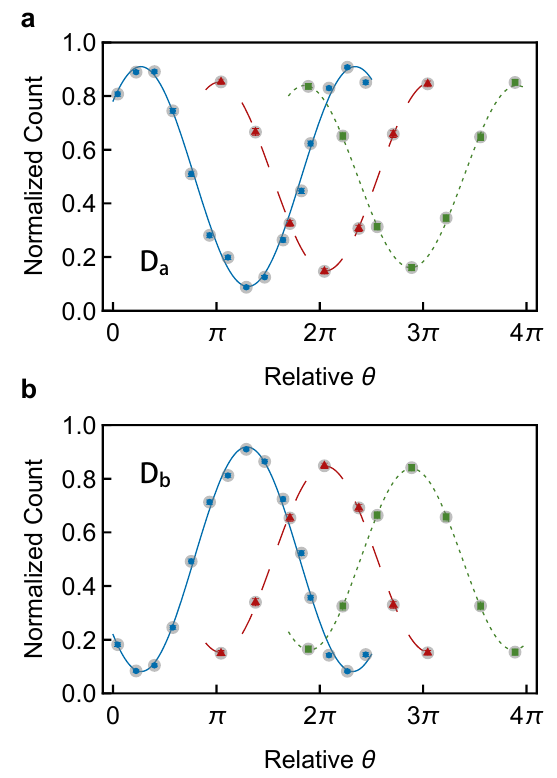}
	\caption{\textbf{Characterization of the remote entanglement via SPI.} When the atomic modes are retrieved as optical modes for interference, the normalized photon count in one output mode of the fiber BS oscillates as a function of the relative phase $\theta$ between the two optical modes. $D_a$ heralded events are shown in \textbf{a}, $D_b$ heralded events are shown in \textbf{b}. Blue squares, red triangles and green dots refer to $L$=10~m, 10~km and 50~km separately. Sinusoids with corresponding color (solid, dashed and dotted in shape) show the fitting results. The result of 50~km is based on $1.7\times10^{5}$ heralding events during a total measurement time of 6 hours over a period of 2 days. The error bars represent one standard deviation.}
	\label{fig:dlcz}
\end{figure}

To verify the Fock-state atomic entanglement, we follow a protocol introduced in Ref~\cite{Chou2005}. The degree of entanglement is quantified in terms of concurrence $\mathcal{C}$, which is a monotone function of entanglement and goes from 0 for a separable state to 1 for a maximally entangled state. Its definition is
$\mathcal{C}=\mathrm{max}(0,2|d|-2\sqrt{p_{00}p_{11}})/P$, where $p_{ij}$ is the probability of having $i$ excitations in ensemble A and having $j$ excitations in ensemble B, $P=p_{00}+p_{01}+p_{10}+p_{11}$, $d\approx V(p_{01}+p_{10})/2$, and $V$ is the interference visibility of the single-excitation states. The excitation statistics of $p_{ij}$ can be measured directly via photon counting of the two read-out modes and applying loss calibration. To measure the interference visibility $V$, we add a relative phase $\theta$ between two read-out modes and mix them via a BS. Along with the scan of $\theta$, counts in two output modes vary as a sinusoidal function of $\theta$ as shown in Fig.~\ref{fig:dlcz}, thus $V_{\theta}$ could be deduced.
In short fiber case ($L$=10~m) without QFC, at $\chi=0.015$, we get a concurrence of $\mathcal{C}=0.677\pm 0.012$ for $\ket{\varPsi^{+}}$ and $\mathcal{C}=0.711\pm 0.012$ for $\ket{\varPsi^{-}}$. In this condition, the entangling probability in one trial is $P_{ent}=0.014$, which is the highest probability of heralded remote entanglement creation to the best of our knowledge.

In the long fiber ($L$=10~km and 50~km) cases, we add the QFC module. The phase noise during long fiber transmission fluctuates faster~\cite{Minavr2008} than the capable band of intermittent phase-locking. Hence we additionally insert an auxiliary continuous 1550~nm laser beam to uninterruptedly monitor phase fluctuation and actively stabilize it (see Supplementary Information). Measured results for the read-out photon interference at different fiber lengths are shown in Fig.~\ref{fig:dlcz}. By fitting the sinusoidal oscillations and measuring the excitation statistics, we get a concurrence result of $\mathcal{C}=0.428\pm 0.013$ at $L$=10~km and $\mathcal{C}=0.407\pm 0.008$ at $L$=50~km $\ket{\varPsi^{+}}$. For $\ket{\varPsi^{-}}$, the results are $\mathcal{C}=0.416\pm 0.008$ at $L$=10~km and $\mathcal{C}=0.348\pm 0.011$ at $L$=50~km. Degradation of concurrence in comparison with the case of short fiber without QFC is mainly due to the remaining noise after phase stabilization (see Supplementary Information), which can be significantly improved by optimizing the feedback loop. The measured heralded entangling probability is $P_{ent}=1.57\times10^{-3}$ for 10~km fiber and $P_{ent}=3.85\times10^{-4}$ for 50~km fiber, which correspond to an entanglement creation time of $T_{ent}=$ 32~ms and 0.65~s respectively.

\begin{table*}[htbp]
	\caption{\textbf{Comparison of two-node experiments.}}
	\label{table1}
    \begin{tabular*}{\textwidth}{@{\extracolsep\fill}cccc}\toprule
        \hline
        Experiment	&  TPI (this work)  & SPI (this work) & NV (2015)~\cite{Hensen2015}  \\ \hline
    	Physical separation	& 0.6 m  & 0.6 m & 1.3 km\\			
    	Overall fiber length $L$ & 22 km  & 50 km & 1.7 km\\
        Entanglement probability $P_{ent}$ & $0.73\times10^{-6}$ & $3.85\times10^{-4}$ & $6.4\times10^{-9}$ \\
        Entanglement quality & $\mathcal{F}=0.720\pm0.027$  & $\mathcal{C}=0.378\pm 0.007$  & $\mathcal{F}=0.92\pm0.03$\\
        Entanglement creation time $T_{ent}$ & 150 s & 0.65 s & $1.3\times10^{3}$ s \\
        Quantum link efficiency $\eta_{link}$~\cite{Humphreys2018} & $1.45\times10^{-3}$ & 0.34  & $4.6\times10^{-4}$ \\
        Assumed memory lifetime $\tau_m$ & $0.22$ s~\cite{Yang2016} & $0.22$ s~\cite{Yang2016} & $0.6$ s~\cite{Bar-Gill2013,Humphreys2018}\\
        \hline \\
    \end{tabular*}\\
{\raggedright \small In the long fiber case, the propagation delay results in a maximal repetition rate of $R_{rep}=C/L$, where $C\simeq2\times10^8$ m/s is the speed of light in fiber. Thus the heralded entanglement creation time is estimated as $T_{ent}=(R_{rep}P_{ent})^{-1}$. In the estimation of $\eta_{link}=\tau_m/T_{ent}$, we make use of state-of-the-art lifetime results, as listed in the last row. We chose the 1.3 km NV experiment~\cite{Hensen2015} for comparison, because it is the only two-node experiment before our work that has a fiber length in the kilometer regime. \par}
\label{summary}
\end{table*}

\msection{Discussion and Outlook}\label{sec:conclusion}
We have experimentally demonstrated two feasible ways to entangle two quantum memories via long-distance photon transmission in optical fibers. We summarize key parameters and results in Tab.~\ref{summary}. Even though fiber distance of the SPI experiment is significantly longer than the TPI experiment, the SPI scheme offers a much higher probability of entanglement creation, because merely a single photon passing through half of the whole link is detected. In contrast the TPI scheme requires detection of two photons passing through the whole link. For the extension to physically separated nodes over long distance, the TPI scheme is straightforward, which merely requires photons being indistinguishable. For the extension of the SPI experiment, it requires more efforts since the scheme is phase-sensitive. According to our analysis in the Supplementary Information, the main difficulty is to achieve phase correlation of remote independent control lasers. We have performed a preliminary test with two lasers locked independently with two ultra-stable cavities, which shows that phase correlation can be built and stable for a duration that is long enough for remote entanglement generation (see Supplementary Information for details). Thus it is also feasible to extend our SPI experiment to long-distance separated nodes.

The quantum link efficiency $\eta_{link}$~\cite{Humphreys2018} defined as the ratio of memory lifetime over entanglement generation time is also an important figure of merit for two-node experiments. In our current work, decoherence due to atom motions results in a memory lifetime of $\sim70$~$\upmu$s, which is significantly smaller than the entanglement generation time. While according to our previous work on a very similar setup~\cite{Yang2016}, applying a 3-dimensional optical lattice can improve the lifetime to the regime of sub-second ($\tau_m = 0.22$ s), based on which the quantum link efficiency is estimated to be $\eta_{link}=0.34$ for SPI and $\eta_{link}=2.9\times10^{-3}$ for TPI. For further improvement of $\eta_{link}$, it is crucial to increase the entanglement generation rate by many promising ways. One may use Rydberg blockade to inhibit the high-order excitations during atom-photon entanglement preparation, and make the preparation process deterministic~\cite{Li2013,Li2016}. One can also make use of the multiplexing technique~\cite{Collins2007, Pu2017,Tian2017,Parniak2017} to prepare multiplexed atom-photon entanglement. Shifting the wavelength to telecom C band, optimizing the coupling efficiencies and using better detectors will also increase the remote entanglement rate significantly.

Extending current experiments to long-distance separate nodes, will enable us to perform advanced quantum information tasks over it, such as efficient quantum teleportation over long distance. By incorporating more quantum memories, our experiment may be extended to entangle multiple quantum memories over long distance via multi-photon interference~\cite{Jing2019}. One may also create two pairs of remote atomic entanglement over two sub-links and extend the distance of atomic entanglement via entanglement swapping, following the scheme of quantum repeater~\cite{Briegel1998}. Concatenating this process will extend the distance to the regime that beats direct transmission~\cite{Sangouard2011}.


\newcommand\msubsection[1]{
\vspace{\baselineskip}
\textbf{#1}
\vspace{0.5\baselineskip}
}

\msubsection{Acknowledgment}

This work was supported by National Key R\&D Program of China (2017YFA0303902, 2017YFA0304000), Anhui Initiative in Quantum Information Technologies, National Natural Science Foundation of China, and the Chinese Academy of Sciences. We acknowledge QuantumCTek for providing the field-deployed fibers.

\msubsection{Author Contributions}

X.-H.B. and J.-W.P. conceived the research. Q.Z., X.-H.B. and J.-W.P. designed the experiment. Y.Y., X.-Y.L., B.J., P.-F.S., R.-Z.F., C.-W.Y. and X.-H.B. carried out the experiment with assistance from all other authors. F.M., M.-Y.Z., X.-P.X. and Q.Z. built the QFC module. W.-J.Z., L.-X.Y. and Z.W. fabricated the SNSPDs. Y.Y., Q.Z., X.-H.B. and J.-W.P. analysed the data and wrote the paper with inputs from all other authors.

\msubsection{Correspondence}

Correspondence and requests for materials should be addressed to Q.Z.~(email: qiangzh@\\ustc.edu.cn), X.-H.B.~(email: xhbao@ustc.edu.cn) and J.-W.P.~(email: pan@ustc.edu.cn).

\msubsection{Competing Interests}

The authors declare that they have no competing financial interests.

\msection{Methods}

\textbf{Time sequences.}\ \
Our experiment runs periodically with each period being composed of an atomic loading phase and an entangling phase. Each loading phase takes 18 ms, during which we reload and cool the atoms and perform active phase-locking. In the entangling phase lasting 2~ms, we repeat entangling trials. For the SPI experiment, each trial lasts $5~\upmu$s including $3~\upmu$s for optical pumping and $2~\upmu$s for write and read process. For the TPI experiment, each trial lasts $11~\upmu$s including $3~\upmu$s for optical pumping and $8~\upmu$s for write and read process. The storage duration (relative delay of the read pulse in comparison with the write pulse) is $7~\upmu$s for the TPI experiment and 100 ns for the SPI experiment.

\msection{Data availability statement}

\textbf{Data Availability.}\ \
The data that support the plots within this paper and other findings of this study are available from the corresponding author upon reasonable request.

\clearpage

\definecolor{orange}{cmyk}{0,0.5,1,0}   
\def\openone{\mathds{1}}

\newcolumntype{C}[1]{>{\PreserveBackslash\centering}p{#1}}
\newcolumntype{R}[1]{>{\PreserveBackslash\raggedleft}p{#1}}
\newcolumntype{L}[1]{>{\PreserveBackslash\raggedright}p{#1}}

\renewcommand{\andname}{\ignorespaces}

\setcounter{figure}{0}
\setcounter{table}{0}
\setcounter{equation}{0}

\onecolumngrid

\global\long\def\theequation{S\arabic{equation}}
\global\long\def\thefigure{S\arabic{figure}}
\renewcommand{\thetable}{S\arabic{table}}
\renewcommand{\arraystretch}{0.6}

\normalsize

\msection{SUPPLEMENTARY INFORMATION}

\section{General Information of experimental setups}
\subsection{Energy Level Scheme}\label{sec:para}
\begin{figure}[!hbtp]
	\centering
    \includegraphics[width=\textwidth]{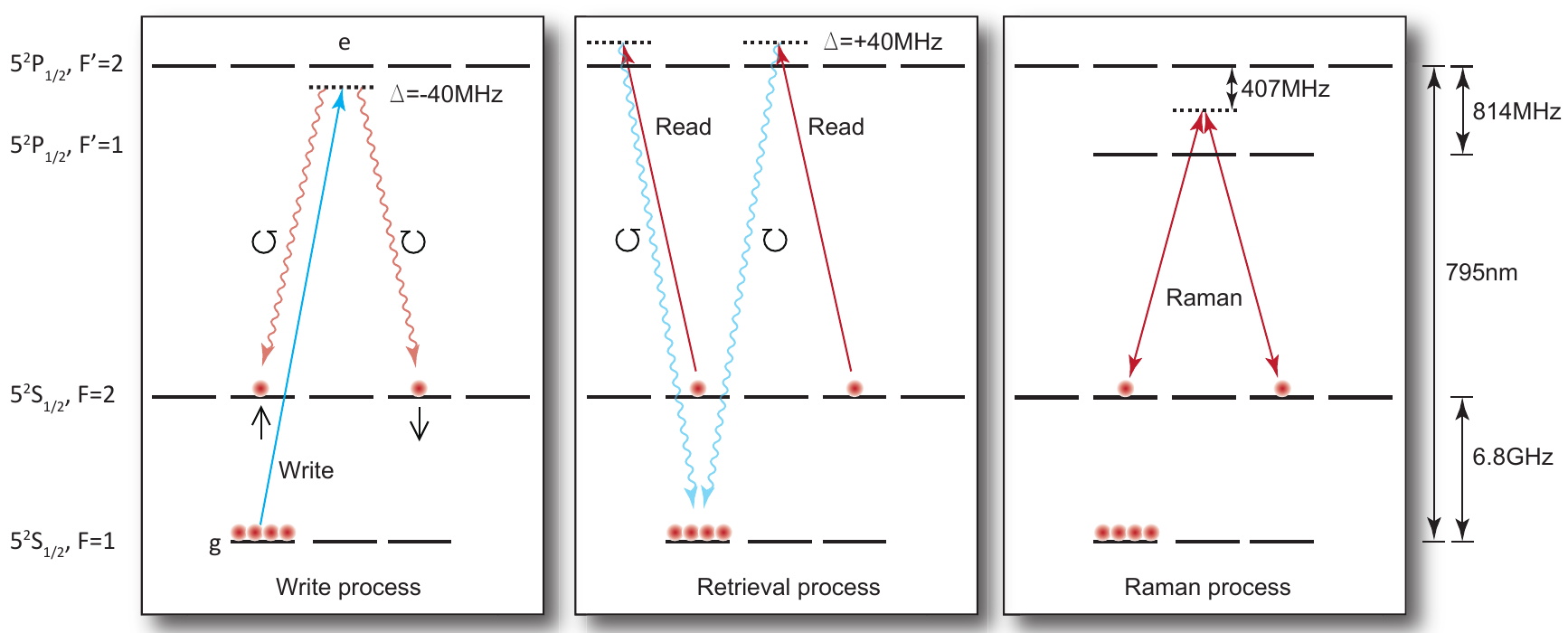}
	\caption{Energy level scheme of the quantum memory.}
	\label{fig:energylevel}
\end{figure}

We use $^{87}Rb$ atoms trapped by the magneto-optical trap as quantum memories. The energy level used in each memory is shown in Fig.~\ref{fig:energylevel}. At the beginning of each experiment, all atoms are prepared in ground state $\ket{g}\equiv\ket{5^2S_{1/2},F=1,m_F=-1}$. Write beam couples $\ket{g}\leftrightarrow\ket{e}\equiv\ket{5^2P_{1/2},F=1,m_F=0}$ with $\Delta=-40$\,MHz. After write process, spin-wave is stored in $\ket{\uparrow}/\ket{\downarrow}\equiv\ket{5^2S_{1/2},F=2,m_F=\mp1}$. Read beam couples $\ket{\uparrow}$ to $\ket{5^2P_{1/2},F=2,m_F=-2}$ and $\ket{\downarrow}$ to $\ket{5^2P_{1/2},F=2,m_F=0}$ both with $\Delta=+40$\,MHz. Ring cavity is resonant with the write-out and read-out photon but not with write or read beam. When performing measurements in $\ket{\uparrow}\pm\ket{\downarrow}$, a Raman pulse couples $\ket{\uparrow}$ ($\ket{\downarrow}$) to $\ket{5^2P_{1/2},F'=1,m_F=0}$ and $\ket{5^2P_{1/2},F'=2,m_F=0}$ simultaneously to perform a $\pi/2$ flip. In the SPI experiment, we merely make use of the optical component $\ket{\circlearrowleft}$  and the atomic component $\ket{\uparrow}$. Memory parameters of lifetime and detected retrieval efficiency are summarized in Tab.~\ref{tab:memory} for both nodes.

\begin{table*}[!hbt]
	\caption{\label{tab:memory}Detailed parameters for the quantum memory. ${\bm\eta_{ret,\uparrow/\downarrow}}$: detected retrieval efficiency of two spin-waves. $\bm{\tau_{m}}$: memory lifetime.}
	\renewcommand\arraystretch{1.3}
	\begin{ruledtabular}
		\begin{tabular}
			{cccc}
			&$\bm{\eta_{ret,\uparrow}}$&$\bm{\eta_{ret,\downarrow}}$&$\bm{\tau_{m}\ (\mu s)}$\\
			\hline
			\textbf{Node A}
			&  $0.333\pm0.011$  &  $0.204\pm0.008$     &$71.19\pm0.09$\\
			\textbf{Node B}
			&  $0.339\pm0.015$  &  $0.218\pm0.010$    &$65.29\pm0.08$ \\
		\end{tabular}
	\end{ruledtabular}
\end{table*}

\subsection{Quantum frequency conversion}
\begin{figure}
	\centering
	
	\includegraphics[width=4.8in]{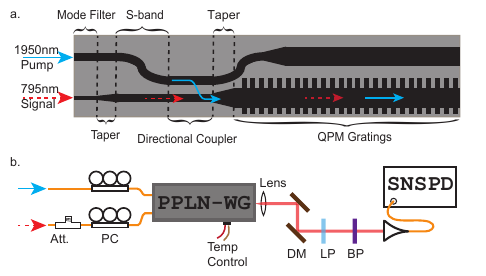}
	
	\caption{ a. Detailed structure inside a PPLN waveguide chip. b. Setup for test of a PPLN waveguides chip. Att.: adjustable attenuation. PC: polarization controller. DM: dichromatic mirror. LP: long pass filter edged at 1150~nm. BP: bandpass filter}
	\label{fig:pplntest}
\end{figure}

We fabricate reverse-proton-exchange (RPE) periodically-poled lithium niobate (PPLN) waveguide chips~\cite{Parameswaran2002,Roussev2004} with a total length of 52~mm for difference-frequency generation (DFG) of 795~nm signal and 1950~nm pump. To couple the two very different wavelengths into the fundamental spatial mode of the same waveguide, we use an integrated waveguide structure consisting of a bent waveguide and a straight waveguide with an entrance center-to-center separation of 126~$\upmu$m, as shown in Fig.~\ref{fig:pplntest}~a.

The main features of the integrated structure are two individual mode filters optimizing the fiber-to-waveguide coupling efficiency of signal and pump respectively, a directional coupler~\cite{Chou1998} working as a wavelength combiner, and a uniform straight waveguide with 45-mm-long QPM gratings for optical frequency nonlinear mixing. The QPM period is 17.1~$\upmu$m.

An HI780 fiber for 795~nm signal and an SM-28e fiber for 1950~nm pump are terminated in a silicon V-groove array and pigtailed to the input of the waveguides. The 795~nm signal is coupled into the 2-$\upmu$m-wide mode filter and the 1950~nm pump is coupled into the 5-$\upmu$m-wide mode filter, with estimated fiber-to-waveguide coupling efficiencies of $89\%$ and $74\%$ respectively. The 1950~nm pump then goes through an adiabatic taper and an S-bend. It finally enters the directional coupler. With a waveguide width of 5.5~$\upmu$m, an edge-to-edge spacing of 3.5~$\upmu$m, and a length of 0.3~mm, the directional coupler combines the 1950~nm pump and the 795~nm signal into the same straight waveguide with low loss simultaneously. We define the pump coupling efficiency as the ratio between the output power of the straight waveguide and the total output power, and define the signal coupling loss as the ratio between the output power of the bent waveguide and the total output power. The measured pump coupling efficiency is -0.3 dB and the signal coupling loss is negligible. The combined waves then enter the nonlinear mixing region with a waveguide width of 7.5~$\upmu$m, where the 795~nm signal is down-converted to 1342~nm telecom O band.

The input and output end faces of the waveguides are anti-reflection (AR) coated for all wavelengths of interest to eliminate the Fresnel reflection loss. After fiber-pigtailing at the input end of the waveguides, the total waveguides throughputs are $70\%$ and $60\%$ for 795~nm and 1950~nm, respectively. A schematic diagram of our experimental setup is shown in Fig.~\ref{fig:pplntest}~b.

A single-frequency fiber laser manufactured by AdValue Photonics is used as the pump source, and 795~nm signal at the single-photon level is attenuated from continuous laser. As RPE lithium niobate waveguides support only TM-polarized modes, polarization controllers (PCs) are used to adjust the polarization for 795~nm and 1950~nm respectively. The working temperature of the waveguides is actively stabilized by a thermoelectric cooling (TEC) system to maintain the phase-matching condition.

The down-converted photons and the remnant pump are collected with an AR-coated aspheric lens, and then the remnant pump is removed with two DMs ($>99.9\%$ reflectivity for 1342~nm and $<5\%$ reflectivity for 1950~nm). A long-pass filter edged at 1150~nm and a band-pass filter centered at 1342~nm with a bandwidth of 5~nm are used in combination to block the noises coming from the strong pump, including the spontaneous Raman scattering (SRS) noise and parasitic noise caused by imperfect periodic poling structure and second and third harmonic generation~\cite{Pelc2011}. By the way, the amplifier spontaneous emission (ASE) from pump laser also contributes a lot around 1342~nm which has been cut off before entering the waveguide. When collecting the 1342~nm DFG signal into single-mode fiber, we achieved around $60\%$ collection efficiency. Totally, the end-to-end efficiency of QFC module is $33\%$.

\section{Phase stabilization}
\subsection{Details of AMZI stabilization in the TPI experiment}

\begin{figure}[!hbtp]
	\centering
	\includegraphics[width=0.6
	\textwidth]{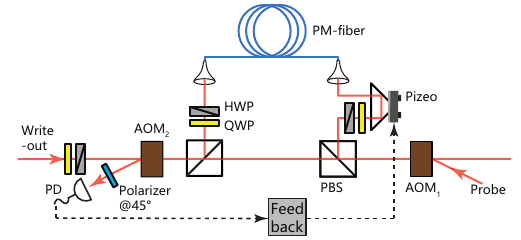}
	\caption{Configuration of AMZI stabilization. AOM: acoustic photonic modulator.}
	\label{fig:amzi}
\end{figure}

\begin{figure}[hbtp]
	\centering
	\includegraphics[width=0.5
	\textwidth]{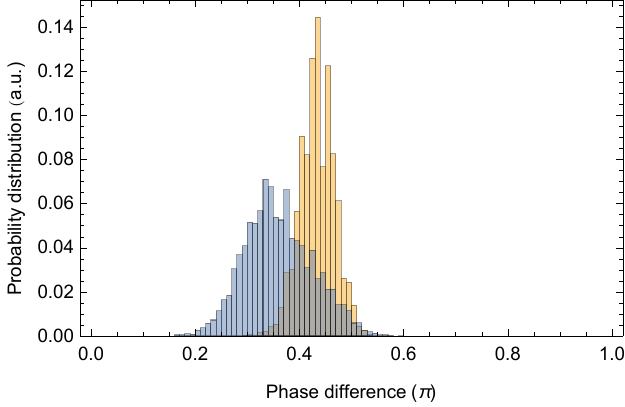}
	\caption{Results of AMZI stabilization. Yellow for probe beam and blue for signal.}
	\label{fig:amziresu}
\end{figure}

In the TPI experiment, to convert polarization photonic qubit to time-bin encoding, we make use of an asymmetric Mach-Zenhder Interferometer (AMZI) in each node. The phase difference between two arms of this interferometer ought to be constant during the whole experiment. To assure this, we perform active phase stabilization as depicted in Fig.~\ref{fig:amzi}. During the experimental phase, two acoustic optical modulators (AOMs) are both switched off. Polarization encoded Write-out photon enters the interferometer and exits as a tim-bin encoded one. In MOT loading phase, we turn both AOMs on. A probe beam with 45$\degree$ linear polarization is led into the AMZI through the diffraction of AOM$_1$. Similarly, it is led out through the diffraction of AOM$_2$ after passing through the AMZI. Evidently, the horizontal and vertical parts of probe beam go through short and long arm, respectively. We make them interfere by a polarizer set at 45$\degree$ and read out the phase information through a fast photo-diode. With the help of a proportional-integral-derivative (PID) circuit and a piezoelectric ceramic, phase variation could be dynamically compensated. Fig.~\ref{fig:amziresu} shows the stabilization performance tested by classical beams.

We note here that, it seems that the idler ports of two PBSs could be used to inject and extract the probe beam. But in this way, the probe beam would propagate across the AMZI in orthogonal polarization with the signal photon. In the long arm, two different polarizations would be coupled to different axises, whose optical paths vary independently.

\subsection{Details of phase-stabilization in the SPI experiment}\label{sec:phase}

\subsubsection{Measurement configuration}\label{subsec:config}

One of the difficulties during measuring $V_{\theta}$ is to assure phase is stabilized between write beams in write process, read beams in retrieval process, pump lasers in QFC process, write-out fields along long fiber channels and read-out fields in measurements. Here we prove that all these problems can be solved by stabilizing two interferometers.

\begin{figure}[hbtp]
	\centering
	\includegraphics[width=0.9
	\textwidth]{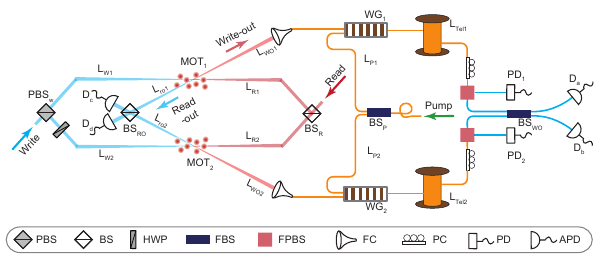}
	\caption{Detailed configuration of $V_{\theta}$ measurement. PBS: polarization beam splitter. BS: beam splitter. HWP: half-wave plate. FBS: fiber beam splitter. FPBS: fiber polarization beam splitter. FC: fiber coupler. PC: polarization controller. PD: photo-diode. APD: avalanche photo-diode.}
	\label{fig:phasesetup}
\end{figure}

As depicted in Fig.~\ref{fig:phasesetup}, we label the distance from $\rm PBS_{W}$ to MOT$_1$ (MOT$_2$) as $L_{W1}$ ($L_{W2}$), distance from $\rm BS_R$ to MOT$_1$ (MOT$_2$) as $L_{R1}$ ($L_{R2}$) and distance from MOT$_1$ (MOT$_2$) to $\rm BS_{RO}$ as $L_{RO1}$ ($L_{RO2}$). Similarly, we denote the distance from MOT$_1$ (MOT$_2$) to the front facet of WG$_1$ (WG$_2$) as $L_{WO1}$ ($L_{WO2}$), the distance from $\rm BS_P$ to the front facet of WG$_1$ (WG$_2$) as $L_{P1}$ ($L_{P2}$), and the distance from the end facet of WG$_1$ (WG$_2$) to $\rm BS_{WO}$ as $L_{Tel1}$ ($L_{Tel2}$). Still, in case the imperfection of manufacturing, we assume the length of two PPLN waveguide chips are different, as $L_{WG1}$ and $L_{WG2}$.

First, we observe the phase evolution in entanglement creation process. Starting from write beam being split in $\rm PBS_W$, we can write the phase of two write beams at two atomic ensembles as:
\begin{equation}\label{eq:writeout1}
\phi_{Wi}=\frac{L_{Wi}}{\lambda_{W}},\ \ \ i=1,2.
\end{equation}
$\lambda_W$ is the wavelength of write beams. Subsequently, they interact with atomic ensembles and Raman scattering happens. Afterwards, the phase of write-out fields and atoms evolve independently. We can write the phase of two write-out fields at the front facets of two waveguides as:
\begin{equation}\label{eq:writeout2}
\phi_{WOi}=\frac{L_{Wi}}{\lambda_{W}}-\phi_{MOTi}+\frac{L_{WOi}}{\lambda_{WO}},\ \ \ i=1,2.
\end{equation}
$\lambda_{WO}$ represents the wavelength of write-out fields and $\phi_{MOTi}$ refers to a time-dependent phase of atoms. Next, in consideration of write-out field propagating, quantum frequency conversion happens and pump beams' phase will also be introduced. Assuming the conversion process always happens in the front facet of waveguides (proof of plausibility in Sec.~\ref{subsec:uncertainty}), we can write the phase right before the $\rm BS_{WO}$ as:
\begin{equation}\label{eq:writeout3}
\phi_{WOi}=\frac{L_{Wi}}{\lambda_{W}}-\phi_{MOTi}+\frac{L_{WOi}}{\lambda_{WO}}+\frac{L_{WGi}+L_{Teli}}{\lambda_{Tel}}-\frac{L_{Pi}}{\lambda_{P}},\ \ \ i=1,2.
\end{equation}
$\lambda_{Tel}$ and $\lambda_{P}$ represent the wavelength of telecom field after conversion and pump beam, respectively.

\begin{figure}[!hbtp]
	\centering
	\
	\includegraphics[width=0.9
	\textwidth]{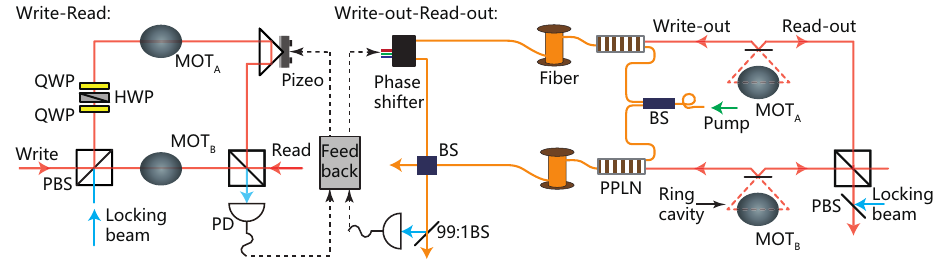}
	\caption{Configuration of phase stabilization. Phase stabilization include two interferometers, i.e. write-read and write-out-read-out.}
	\label{fig:phaseconfig}
\end{figure}

Similarly, we could write the phase of two read-out fields before the $\rm BS_{RO}$ as:
\begin{equation}\label{eq:readout}
\phi_{ROi}=\frac{L_{Ri}}{\lambda_{R}}+{\phi_{MOTi}}'+\frac{L_{ROi}}{\lambda_{RO}},\ \ \ i=1,2.
\end{equation}
The difference between $\phi_{MOTi}$ and ${\phi_{MOTi}}'$ comes from phase evolution of a certain energy level of atom. Because the interval time between write and read process is fixed, it is obvious that
\begin{equation}\label{eq:mot}
{\phi_{MOTi}}'-\phi_{MOTi}=constant.
\end{equation}
The phase condition to be fulfilled is:
\begin{equation}\label{eq:interferometer}
\phi_{W1}+\phi_{R1}=\phi_{W2}+\phi_{R2}+2n\pi,
\end{equation}
where $n$ is an integer. We find a solution to Eq.~\ref{eq:interferometer} as below,
\begin{equation}\label{eq:inter1}
\frac{L_{W1}}{\lambda_{W}}+\frac{L_{R1}}{\lambda_{R}}=\frac{L_{W2}}{\lambda_{W}}+\frac{L_{R2}}{\lambda_{R}}+2n'\pi
\end{equation}
\begin{equation}\label{eq:inter2}
\begin{split}
\frac{L_{RO1}}{\lambda_{RO}}+&\frac{L_{WO1}}{\lambda_{WO}}+\frac{L_{WG1}+L_{Tel1}}{\lambda_{Tel}}-\frac{L_{P1}}{\lambda_{P}}\\
&=\frac{L_{RO2}}{\lambda_{RO}}+\frac{L_{WO2}}{\lambda_{WO}}+\frac{L_{WG2}+L_{Tel2}}{\lambda_{Tel}}-\frac{L_{WG2}}{\lambda_{P}}+2m'\pi
\end{split}
\end{equation}
where $n'$ and $m'$ refer to two integers. Eq.~\ref{eq:inter1} and Eq.~\ref{eq:inter2} refer to two Mach-Zehnder interferometers as depicted in Fig.~\ref{fig:phaseconfig}. The first one takes $\rm PBS_{W}$ and $\rm BS_{R}$ as two beamsplitters of the interferometer and covers paths of write and read beams. Therefore we introduce a 795~nm locking beam from the idle port of Read-BS to detect the interference signal in the second output of the $\rm BS_W$ and feed it back to a piezoelectric ceramic. A combination of two half-wave-plate (HWPs) and a quarter-wave-plate (QWP) in the sandwich configuration introduce a relative phase $\theta$ between phase-locking laser and the write beam without changing their polarization.

The second interferometer takes $\rm BS_{WO}$ and $\rm BS_{RO}$ as two beamsplitters and covers paths of write-out and read-out photons, meanwhile includes the frequency conversion modules and two several kilometer long fiber coils. We introduce another locking beam from $\rm BS_{RO}$ using an unbalanced BS ($\frac{Reflectivity}{Transmission}=99:1$) and detect the interference signal in the $\rm BS_{WO}$ in the same way. The frequency of this beam is far detuned from the resonance point of the cavity, thus it neither enters the cavity nor has any interaction with the atoms. In this method, we assume the zero scale of our atomic ensemble. Actually, there still exist a little uncertainty of phase difference introduced by the non-zero scale of ensemble. But we will prove that it is small enough in Sec.~\ref{subsec:uncertainty}.

\subsubsection{Suppression of fast phase variation in long fiber situation}
\begin{figure}[hbtp]
	\centering
	\
	\includegraphics[width=0.6
	\textwidth]{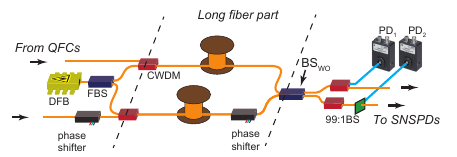}
	\caption{Configuration of assistant phase stabilization in long fiber situation.}
	\label{fig:longfiber}
\end{figure}

Although the phase stabilization method described in Sec.~\ref{subsec:config} works well in short fiber situation, it cannot sustain a good stabilization in long fiber situation. This is because in our former stabilization protocol, phase stabilization only works in MOT loading phases but is turned off during each 2~ms experimental phase. Hence the phase actually randomly fluctuates during these periods. The fluctuation is small enough in short fiber case but too big to be accepted in long fiber case. Therefore, we introduce an extra phase stabilization for long fiber part as depicted in Fig.~\ref{fig:longfiber}. A 1550~nm laser beam is lead in right behind the QFC modules, and lead out after $\rm BS_{WO}$ with the help of coarse wave demultiplexing modulers (CWDMs). Because its wavelength is far enough from our 1342~nm signal in the spectrum, this 1550~nm beam and the stabilization process could run continuously. But, for the same reason, its stabilization reasult does not reprsent the phase we care about. It can only serve as an assistance to help us to suppress the phase fluctuation when phase locking laser is off.

\subsubsection{Phase uncertainty in PPLN waveguide chips and atomic ensembles}\label{subsec:uncertainty}

\begin{figure}
	\centering
	
	\includegraphics[width=4.8in]{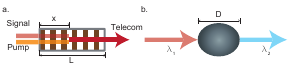}
	
	\caption{a. Transition point uncertainty in a PPLN waveguides chip. b. Exciting position uncertainty in MOT. }
	\label{fig:uncertainty}
\end{figure}

In DFG process, we always have the energy conservation principal
\begin{equation}
\frac{1}{\lambda_{Signal}}=\frac{1}{\lambda_{Pump}}+\frac{1}{\lambda_{Telecom}}.
\end{equation}
Supposing DFG happens $x$ from the front facet of the PPLN waveguide chip as depicted in Fig.~\ref{fig:uncertainty}.a, then the phase accumulation of optical field is
\begin{equation}
\begin{aligned}
\phi=&\frac{x}{\lambda_{Signal}}-\frac{x}{\lambda_{Pump}}+\frac{L-x}{\lambda_{Telecom}}\\
=& x\cdot(\frac{1}{\lambda_{Pump}}+\frac{1}{\lambda_{Telecom}})-\frac{x}{\lambda_{Pump}}+\frac{L-x}{\lambda_{Telecom}}\\
=&\frac{L}{\lambda_{Telecom}}.
\end{aligned}
\end{equation}
which means we can always assume DFG happening at the front facet of the PPLN waveguides chip.

In the former discussion about interferometer such as Eq.~\ref{eq:inter1} and Eq.~\ref{eq:inter2}, we assume the atomic ensemble as a point with no scale and was settled in a certain point. Then we can ignore the change of the wavelength after this point and using one laser to stabilize phase. Now we consider a real ensemble with diameter $D$ as depicted in Fig.~\ref{fig:uncertainty}.b, which is around $100~\upmu$m in our system. Two lasers here are both around $\lambda=795$~nm with $\delta=6.8$~GHz difference. First, we assume the location where spontaneous Raman scattering happens in the left and right edge of the ensemble. The phase difference is
\begin{equation}
\begin{aligned}
\Delta\theta=& 2\pi\cdot(D/\lambda_1-D/\lambda_2)\\
=& 2\pi D\cdot\frac{\delta}{c}\approx0.014=0.81^{\circ}.
\end{aligned}
\end{equation}
It is plausible to consider this location obeys uniform distribution in this regime, we can easily know the standard deviation of phase difference $S$ as
\begin{equation}
S=\sqrt{\frac{{\Delta\theta}^2}{12}}\approx4\times10^{-3}=0.24^{\circ}.
\end{equation}
So, theoretically, in the perfect stabilizing condition, only $0.24^{\circ}$ uncertainty of phase will be introduced.

\section{Lasers in outdoor application}\label{sec:laserphase}
In current experiment, two atomic ensembles are placed nearby in one laboratory. For simplicity, two ensembles share control beams (i.e. write, read and pump beams) from same lasers. Moreover, in phase stabilization process of the SPI experiment, auxiliary beams to detect the phase difference between two paths are also split from one laser. In TPI experiments, lasers located in different nodes could be easily locked to an absolute frequency standard such as absorption spectrum or ultra-stable cavity. But in SPI experiments, not only the frequency, but also the phase of lasers in different nodes need to be synchronized. There must be a question that is it necessary to share these control and auxiliary beams among distant nodes or actively lock their relative phase in outdoor application. Here we propose a practical protocol to deal with phase fluctuation in outdoor application. Furthermore, we did a brief experimental test to simulate some scenarios in future experiments. The result supports the feasibility of our protocol.

\begin{figure}[hbtp]
	\centering
	\includegraphics[width=0.8
	\textwidth]{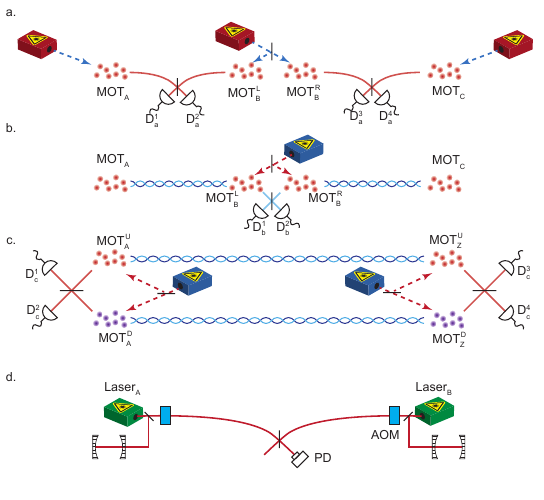}
	\caption{Three main steps in DLCZ protocol. a. Heralded entanglement creation in basic segments. b. Entanglement swapping. c. Converting two EME states to a PME state. d. Test of phase fluctuation between two independent lasers.}
	\label{fig:laserset}
\end{figure}

\subsection{Phase of control lasers}
In recently developed twin-field quantum key distribution (TF-QKD)\cite{lucamarini2018}, single photon interference is also utilized, which brings similar questions about phase. Based on their solution\cite{liu2019}, we propose a practical protocol to cancel phase fluctuation and uncertainty in SPI outdoor experiment.

There are three main steps in DLCZ protocol, heralded entanglement creation in basic segments, a series of entanglement swappings and finally converting two effective maximally entangled (EME) states to a polarization maximally entangled (PME) state. Without loss of generality, we observe three normal cases here to represent three steps respectively.

First, in entanglement creation step, we put only one laser in each node serving as write beam as depicted in Fig.~\ref{fig:laserset}~a. In node-B, write beams for two ensembles are split from one laser. Suppose each laser has an initial phase $\phi_A$, $\phi_B$ and $\phi_C$ respectively. Besides, we assume that all lasers work in the same frequency $\omega_w$. When $D_a^1$ and $D_a^3$ click, the state of four ensembles becomes
\begin{equation}
\begin{split}
|\Psi\rangle_{AB_LB_RC}=&(e^{i\phi_A}|1_S0_S\rangle_{AB_L}+e^{i\phi_B}e^{i\theta_{AB}}|0_S1_S\rangle_{AB_L})\\&\otimes
(e^{i\phi_B}|1_S0_S\rangle_{B_RC}+e^{i\phi_C}e^{i\theta_{BC}}|0_S1_S\rangle_{B_RC})e^{i\omega_w\Delta t},
\end{split}
\end{equation}
where $\Delta t$ is time difference between entanglement $MOT_A-MOT_B^L$ and $MOT_B^R-MOT_C$ creation. $\theta_{AB}$ and $\theta_{BC}$ are phase of the long fibers. Besides, the pump beam in QFC process brings an extra phase (initial phase of pump laser) between $|0_S\rangle$ and $|1_S\rangle$, which is similar as the write beam. Therefore we consider the pump laser as a part of write laser and do not list it in the equation.

Here, the first step of our protocol is that, right before or after each trial of entanglement creation, every memory node sends a series of strong reference pulses to the middle station to estimate phase difference between two nodes, for instance, $\phi_{AB}=\phi_B+\theta_{AB}-\phi_A$ for node A and B and $\phi_{BC}=\phi_C+\theta_{BC}-\phi_B$ for node B and C. Then the second step is to send the estimation results back to memory nodes.

Next we perform entanglement swapping as in Fig.~\ref{fig:laserset}~b. Spin-waves in $MOT_B^L$ and $MOT_B^R$ are mapped onto read-out optical fields with the help of read beams in frequency $\omega_r$. Since they are placed in one node, read beams are split from one laser with the initial phase $\psi_B$. After retrieval, the state of two ensembles and two optical fields becomes
\begin{equation}
\begin{split}
|\Psi\rangle_{AB_LB_RC}=&(e^{i\phi_A}|1_S0_{ro}\rangle_{AB_L}+e^{i(\phi_B+\psi_B)}e^{i\theta_{AB}}|0_S1_{ro}\rangle_{AB_L})\\&\otimes(e^{i(\phi_B+\psi_B)}|1_{ro}0_S\rangle_{B_RC}+e^{i\phi_C}e^{i\theta_{BC}}|0_{ro}1_S\rangle_{B_RC})e^{i\omega_w\Delta t}.
\end{split}
\end{equation}
After two read-out fields interfere at middle BS, entanglement is swapped to $MOT_A$ and $MOT_C$ as
\begin{equation}
|\Psi\rangle_{AC}=(e^{i\phi_A}|1_S0_S\rangle_{AC}+e^{i\phi_C}e^{i(\theta_{AB}-\theta_{BC})}|0_S1_S\rangle_{AC})e^{i\omega_w\Delta t}.
\end{equation}
Generally, during each quantum swapping process, the initial phase from middle laser is eliminated and the phase of final state only comes from the long fiber and laser in end nodes.

Finally, through entanglement creation in basic segments and a series of swapping, we get a pair of entanglements $MOT_A^U-MOT_Z^U$ and $MOT_A^D-MOT_Z^D$. We convert these two EME states to a PME state. Mapping all spin-waves onto read-out fields, we get
\begin{equation}
\begin{split}
|\Psi\rangle_{A_UZ_UA_DZ_D}=&(e^{i(\phi_A+\psi_A)}|1_{ro}0_{ro}\rangle_{A_UZ_U}+e^{i(\phi_Z+\psi_Z)}e^{i\theta_{AZ}}|0_{ro}1_{ro}\rangle_{A_UZ_U})\\&\otimes(e^{i(\phi_A'+\psi_A)}|1_{ro}0_{ro}\rangle_{A_DZ_D}+e^{i(\phi_Z'+\psi_Z)}e^{i\theta_{AZ}'}|0_{ro}1_{ro}\rangle_{A_DZ_D})e^{i\omega_w\Delta t}e^{i\omega_r\Delta t'},
\end{split}
\end{equation}
where $\phi_A$ ($\psi_A$) and $\phi_Z$ ($\psi_Z$) refer to initial phase of write (read) laser in node-A and Z. The prime on some terms indicates a discrepancy from prime-free one, arising from too much interval time larger than coherence time of the laser, or too fast fiber fluctuation. $\Delta t'$ is the time difference between entanglement $MOT_A^U-MOT_Z^U$ and $MOT_A^D-MOT_Z^D$ generation. When we register only the coincidences of two-side detectors, the effective part contributing to the final result is

\begin{equation}\label{eq:finalstate}
\begin{split}
|\Psi\rangle_{AZ}&=(|1_{ro}0_{ro}\rangle_{AZ}+e^{i(\phi_Z+\theta_{AZ}-\phi_A)}e^{i(\phi_Z'+\theta_{AZ}'-\phi_A')}|0_{ro}1_{ro}\rangle_{AZ})e^{i\omega_w\Delta t}e^{i\omega_r\Delta t'}.\\
&=(|1_{ro}0_{ro}\rangle_{AZ}+e^{i(\phi_{AZ}+\phi_{AZ}')}|0_{ro}1_{ro}\rangle_{AZ})e^{i\omega_w\Delta t}e^{i\omega_r\Delta t'}.
\end{split}
\end{equation}
The last step of our protocol is to calculate $\phi_{AZ}$ and $\phi_{AZ}'$ from $\phi_{AB}$, $\phi_{AB}'$, $\phi_{BC}$, ... and compensate it to state in Eq.~\ref{eq:finalstate} via a phase modulator.

We can see that only the first step is challenging. It requires a fast estimation of phase and a slow phase variation to meet the consistency of phase between detection and experiment moments. In our situation, the interval time between two trials is about 5~$\upmu$s. Supposing detecting phase difference 2~$\upmu$s before each trial, we need an estimation fast enough, which is already achieved in ref.~\cite{liu2019}, and a phase variation slow enough in this time scale.

\subsection{Test result}
\begin{figure}[!hbtp]
	\centering
	\includegraphics[width=1
	\textwidth]{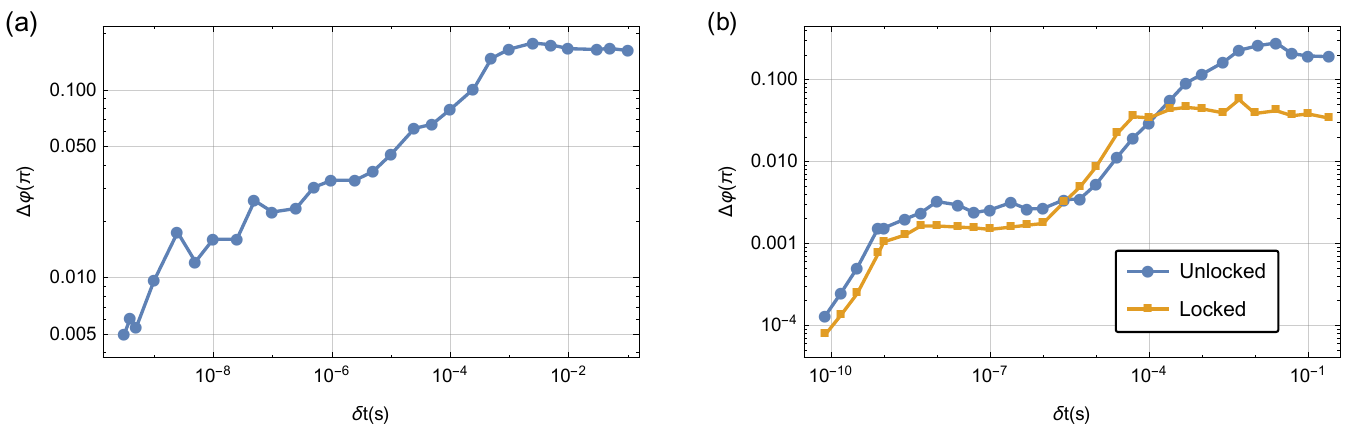}
	\caption{a. Deviation of phase difference between two independent lasers in different sampling interval. b. Deviation of phase difference between two long fibers in different sampling intervals.}
	\label{fig:allantest}
\end{figure}

Based on the existing condition of our lab, we perform two tests to simulate the situation when two nodes are physically separated. First, we test the phase stability between two independent lasers as shown in Fig.~\ref{fig:laserset}. We lock two 795~nm lasers (TOPTICA DLpro) to two ultra-stable cavities (Stable Laser System ATF6010-4) separately. Their locking points are around 170~MHz apart and linewidths are less than 5~kHz. After shifting their frequency together with two acoustic optical modulators, we interfere two lasers with the help of a beamsplitter. Then we detect and record its result via a high-speed photodiode (Thorlabs PDA8GS; 9.5~GHz Bandwidth) and an oscilloscope (Keysight DSO-X 4054A; 500~MHz Bandwidth, 5~GHz sampling rate) as depicted in Fig.~\ref{fig:laserset}.d. Choosing different sampling intervals $\delta t$, we can record the phase fluctuation $\Delta\varphi=\varphi(t+\delta t)-\varphi(t)$ as shown in Fig.~\ref{fig:allantest}.a of phase fluctuation (similar to Allan deviation in frequency). We can see that in the time scale we care about ($\delta t=$2~$\upmu$s), its value is $0.03\pi$. This means if we detect the relative phase between two lasers in two nodes as $\varphi_0$, 2~$\upmu$s before each trial of the single-photon-interference experiment, the phase actually contributes to the entangled state is $\varphi=\varphi_0 \pm 0.03\pi\approx \varphi_0\pm5.4\degree$, which is totally acceptable. And, we can further suppress this value by performing this detecting before and after each trial simultaneously.

Next, we observe the stability of fibers. We did similar statistics to 50~km fiber interferometer in SPI experiment. Due to two arms sharing one laser, only fiber fluctuation contributes to this result. In Fig.~\ref{fig:allantest}.b, we can see that when $\delta t=$2~$\upmu$s, $\Delta\varphi<0.005\pi$ both in locking and unlocking case, which is negligible compared with laser phase fluctuation.

\subsection{Statistics of phase fluctuation}\label{sec:phaseallan}

In the phase measurement, we need to infer the phase from intensity, which is directly detected. However, the mapping from phase to intensity is not one-to-one correspondence but periodical. This may bring a few problem about this measurement. First, one cannot infer an absolute phase. But this is not critical because only the phase difference between two adjacent sampling points is important. Second, obviously, the larger the phase difference, the bigger odds it is misrecognized. Phase difference larger than $\pi$ would certainly be recognized as a small one, for instance. Third, near the extremum of intensity fringe, i.e. phase around $n\pi$, even a small phase difference would lead to misjudgment.

\begin{figure}[hbtp]
	\centering
    \includegraphics[width=	0.5\textwidth]{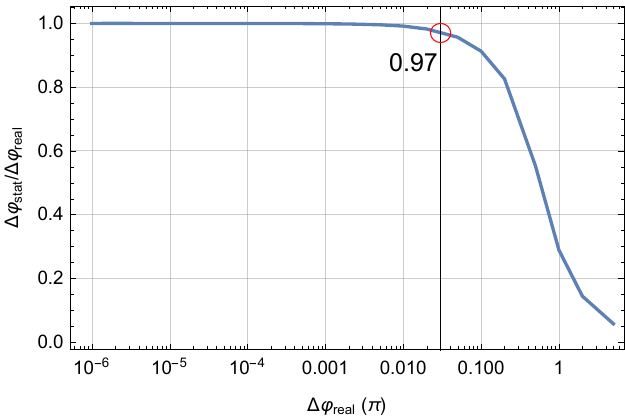}
	\caption{Simulation result of test accuracy $A\equiv\Delta_{stat}/\Delta_{real}$ along with the variation of $\Delta_{real}$.}
	\label{fig:allansim}
\end{figure}

Second and third question may lead to underestimation of real result. To ensure a trustworthy result, we briefly analyze the influence of above questions. Supposing the phase at moment t is $\varphi(t)$, it would be distributed evenly in phase space. After a period $\delta t$ has passed, due to random fluctuation, it is predictable that the phase in $t+\delta t$ is obeys an Gaussian distribution $N(\varphi(t),\Delta\varphi)$ with standard deviation $\Delta\varphi$ centered at $\varphi(t)$. We simulate this process by random sampling in computer. In given phase $\Delta\varphi_{real}$, we first randomly choose $\varphi(t)$ in phase space and give a series of phase points obeying $N(\varphi(t),\Delta\varphi_{real})$. Then we calculate the intensity of each phase and derive it back to a phase, which is possibly different from the original one. Finally do statistics of this phase and average it for all $\varphi(t)$ to give a statistical result $\Delta\varphi_{stat}$. By defining test accuracy as $A\equiv\Delta_{stat}/\Delta_{real}$, we show the simulation result in Fig.~\ref{fig:allansim}. Apparently, in Fig.~\ref{fig:allantest}, results after $\Delta\varphi>0.1$ are untrustworthy. For $\Delta\varphi\leq 0.03\pi$, accuracy of test result is larger than 0.97.

\section{Analysis on experimental imperfections}

\subsection{TPI experiment}
\subsubsection{Imperfection of photon interference}
\begin{figure}
	\centering
	
	\includegraphics[width=1\textwidth]{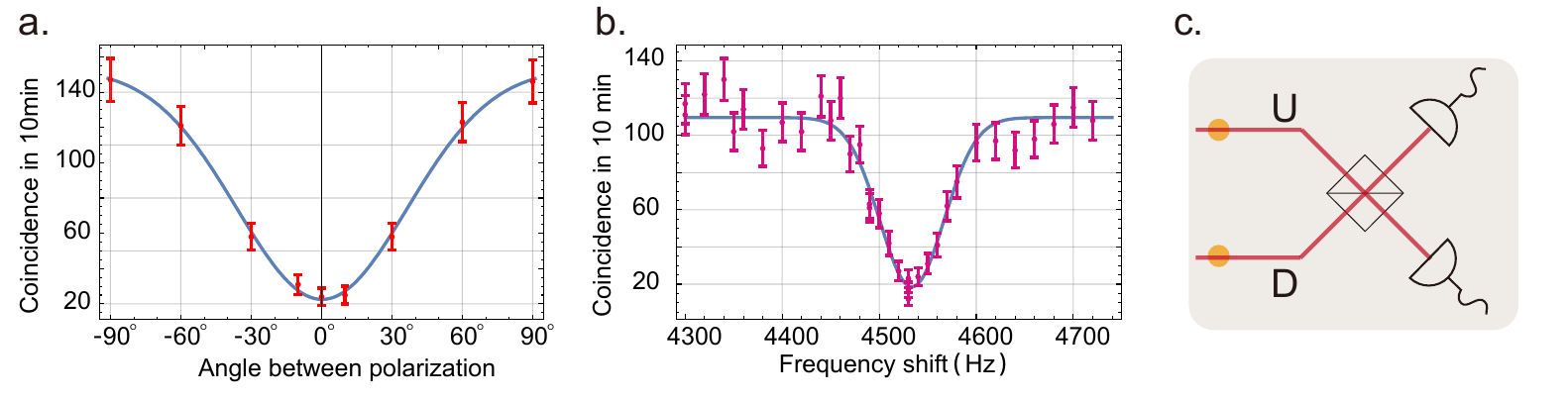}
	
	\caption{Homogeneity of photons from different ensembles. a and b. Hong-Ou-Mandel results of write-out and read-out photon. c. Experimental setup.}
	\label{fig:hom}
	
\end{figure}

Photon interference lies in the very heart of remote entanglement generation, which requires a good indistinguishability of photons from different ensembles. Hence we check it via Hong-Ou-Mandel (HOM)~\cite{Hong1987} experiment. As depicted in Fig.~\ref{fig:hom}.a, we get HOM visibility $V\equiv\frac{N_{min}}{2N_{average}}$ as 0.082 for write-out photon from two ensembles. Based on this, we calculate its influence on experiments. In the HOM experiment, two photons interference at the BS as:
\begin{equation}
a_U^{\dagger}b_D^{\dagger}\rightarrow(a_U^{\dagger}+ia_D^{\dagger})(b_D^{\dagger}+ib_U^{\dagger}),
\end{equation}
where $a^{\dagger}$ and $b^{\dagger}$ are creation operator of photons from different ensembles. Subscript denote their path information. We write the creation operator $b^{\dagger}$ as the superposition of $a^{\dagger}$ and $\tilde{a}^{\dagger}$ (in the orthogonal space of $a^{\dagger}$) as $b^{\dagger}=\alpha a^{\dagger}+ \beta \tilde{a}^{\dagger}$. Then we get
\begin{equation}
a_U^{\dagger}b_D^{\dagger}\rightarrow\alpha(a_U^{\dagger}+ia_D^{\dagger})(a_D^{\dagger}+ia_U^{\dagger})+\beta(a_U^{\dagger}+ia_D^{\dagger})(\tilde{a}_D^{\dagger}+i\tilde{a}_U^{\dagger}).
\end{equation}
Term with coefficient $\alpha$ represents perfect two-photon interference and contributes no coincidence. Term with coefficient $\beta$ represents no interference and contributes coincidence with probability $1/2$. Thus we can write the HOM visibility as $V_{HOM}=\frac{1}{2}\beta^2$.

For the TPI experiment, we characterize the inhomogeneity by considering an imperfect two-photon interference, i.e. BSM measurement, towards write-out photon as
\begin{equation}
\hat{S}^{\pm}=(1-\lambda)\ket{\Psi^{\pm}_{pp}}\bra{\Psi^{\pm}_{pp}}+\lambda\ket{\Psi^{\mp}_{pp}}\bra{\Psi^{\mp}_{pp}}.
\end{equation}
$\lambda$ is a parameter to evaluate the perfectness of BSM and $\ket{\Psi_{pp}^{\pm}}\equiv(\ket{EL}\pm\ket{LE})/\sqrt{2}$ are two of four photon-photon Bell states. The first term denotes an ideal BSM, and the second term denotes the BSM giving the wrong output. In our configuration, there's no possibility of projecting state onto the other two Bell states. Here we build its relation with HOM result. Given two photons entangles perfectly as $\ket{\Psi^{\pm}_{pp}}$, we can write their state by creation operators as
\begin{equation}
\ket{\Psi^{\pm}_{pp}}=a^{\dagger}_Eb^{\dagger}_L\pm a^{\dagger}_Lb^{\dagger}_E\ket{vac} =\Big(\alpha(a^{\dagger}_Ea^{\dagger}_L+a^{\dagger}_La^{\dagger}_E)\pm\beta(a^{\dagger}_E\tilde{a}^{\dagger}_L+a^{\dagger}_L\tilde{a}^{\dagger}_E)\Big)\ket{vac}.
\end{equation}
The $E$ and $L$ in subscript denote early and late time-bin mode, respectively. When they are interacting in a BS as in HOM experiment, the first term will lead to identical photon interference, i.e. a perfect BSM. On the contrary, the second term will lead to presence of two distinguishable photons, which gives a random coincidence result. This means a perfect entangled state get chance of $\frac{1}{2}\beta^2$ being mis-recognized. Following this logic, we can get $\lambda=\frac{1}{2}\beta^2$. Therefore, joint state $\rho_{aa}^{AB}$ between two atomic ensembles after BSM and its fidelity could be expressed as\cite{nielsen2002}:
\begin{gather}
\rho_{aa}^{AB}=\frac{Tr_{pp}[(\openone_{aa}\otimes\hat{S}^{\pm})\rho_{ap}^A\rho_{ap}^B]}{Tr[(\openone_{aa}\otimes\hat{S}^{\pm})\rho_{ap}^A\rho_{ap}^B]},\\
\mathcal{F}=Tr[(\ket{\Psi^{\pm}_{aa}}\bra{\Psi^{\pm}_{aa}})\rho_{aa}^{AB}].\label{eq:fid}
\end{gather}
$\rho^{A}_{ap}$ and $\rho^{B}_{ap}$ are atom-photon entanglement in two nodes. $a$ and $p$ in script refer to atoms and photon, respectively. Note that the HOM result is also influenced by multiple photons, which is already counted when we perform the tomography to $\rho_{ap}^A$ and $\rho_{ap}^B$. So we amend HOM visibility by reducing its influence and get $V'_{wo}=0.063$. We can calculate the result of \ref{eq:fid} and get $\mathcal{F}=0.835$, which is similar with experimental result
$0.798\pm0.063$ and $\mathcal{F}=0.829\pm0.036$ for $\ket{\varPsi^+}$ and $\ket{\varPsi^-}$ in the local case.
\subsubsection{Subtraction of accidental coincidence}
In TPI experiment, the expected four-body coincidence should be given by two write-out photons from two memories interfering and being detected in the middle and two correlated read-out photons being detected on two sides, which shows up in probability $P_{exp}$ as:
\begin{equation}
	P_{exp}=0.5p_w^2\eta_{ret}^2,
\end{equation}
where $p_w$ is the detected probability of write-out photon, equaling to $\chi$ multiplying efficiency of optical components and detector. $\eta_{ret}$ is the detected retrieval efficiency of read-out photon. Nevertheless, our quantum memory is based on spontaneous Raman scattering. Similar to spontaneous parametric down-conversion (SPDC) process, accidental coincidences emerge from simultaneous detection of uncorrelated photons\cite{tittel1998}. Here we consider two main sources of accidental coincidences. First is that after a successful BSM, one of read-out photon is missed for inefficiency but an uncorrelated photon in its mode is detected. Second is that when two write-out photons from one memory make a coincidence in the BSM, an uncorrelated photon in the other memory's read-out mode is detected. These two terms could be estimated as:
\begin{equation}
P_{acc}=0.5p_w^2\eta_{ret}(1-\eta_{ret})p_r\times2+0.5(\frac{p_w}{\sqrt{2}})^2[2\eta_{ret}(1-\eta_{ret})]p_r\times2,
\end{equation}
where $p_r$ is the detected probability of read-out photon. Due to randomness, the accidental coincidences will contribute to parallel and cross correlation equally. After subtraction of them, $\mathcal{F}$ basically keep invariant along $\chi$ near unity taking error bar into consideration.

\subsection{SPI experiment}
\subsubsection{Imperfection of photon interference}
Similar to the TPI case, we consider a imperfect single photon interference process. Because the write-out and read-out field both interfere, we express them as two operators:
\begin{equation}
\begin{split}
	\hat{S}^{\pm}_{wo}=\lambda\ket{\Psi^{\pm}}\bra{\Psi^{\pm}}+(1-\lambda)\ket{\Psi^{\mp}}\bra{\Psi^{\mp}},\\
	\hat{S}^{\pm}_{ro}=\lambda'\ket{\Psi^{\pm}}\bra{\Psi^{\pm}}+(1-\lambda')\ket{\Psi^{\mp}}\bra{\Psi^{\mp}}.
\end{split}
\end{equation}
It is easy to get $\lambda=V_{wo}$ and $\lambda'=V_{ro}=0.074$ similar to the TPI case. $V_{\theta}$ corresponds to the measurement result is:
\begin{equation}
	V_{\theta}=\frac{Tr[(\hat{S}_{ro}^{\pm}-\hat{S}_{ro}^{\mp})\hat{S}_{wo}^{\pm}\rho_A^F\rho_B^F]}
	{Tr[\hat{S}_{wo}^{\pm}\rho_A\rho_B]}.
\end{equation}
$\rho_A^F$ and $\rho_B^{F}$ refer to the Fock state entanglement between write-out and read-out field, which is not known in our experiment. Hence we feed into an ideal maximally entangled state as
\begin{equation}
	\rho_A^F=\rho_B^F=\ket{\Psi^+}\bra{\Psi^+}.
\end{equation}
After we subtract multiple excitations relatedly counted in write-out and read-out HOM, simulation result shows that for this ideal state, $V_{\theta}=0.827$, which is similar to the non-conversion case as listed in Tab.~\ref{tab:res}.

\subsubsection{Phase Instability}
\begin{figure}[!hbtp]
	\centering
	\includegraphics[width=0.5\textwidth]{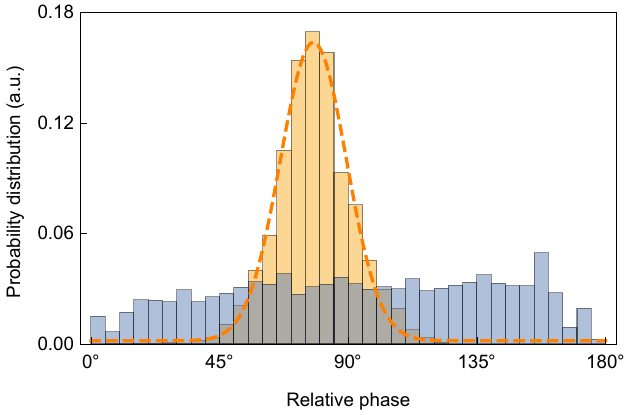}
	\caption{Phase fluctuation of 50~km situation with (yellow) and without (blue) stabilization. Dashed curve is the fitted result.}
	\label{fig:phasedistri}
\end{figure}

In the long fiber situation, we probe the phase instability by monitoring the phase stabilized laser. Through Gaussian fit, we deduce $8.3\degree$ and $13.4\degree$ fluctuation for 10~km and 50~km situation separately. Fig.~\ref{fig:phasedistri} shows the statistic results of 50~km. Regarding phase fluctuation as a small disturbance we consider disturbed state $\rho_{AB}'$, in which $|\Psi^+\rangle$ and $|\Psi^-\rangle$ are transfered to  $|\Psi_{\delta\theta}^+\rangle=|01\rangle+e^{i\delta\theta}|10\rangle$ and $|\Psi_{\delta\theta}^-\rangle=|01\rangle-e^{i\delta\theta}|10\rangle$. Taking phase disturbance account, we have
\begin{equation}
\begin{aligned}
V_{\theta}'=\frac{max'-min'}{max'+min'}&=\frac{tr(|\Psi_{\delta\theta}^+\rangle\langle\Psi_{\delta\theta}^+|\rho_{AB}')-tr(|\Psi_{\delta\theta}^-\rangle\langle\Psi_{\delta\theta}^-|\rho_{AB}')}{tr(|\Psi_{\delta\theta}^+\rangle\langle\Psi_{\delta\theta}^+|\rho_{AB}')+tr(|\Psi_{\delta\theta}^-\rangle\langle\Psi_{\delta\theta}^-|\rho_{AB}')}\\
&=\frac{(p_+-p_-)\cdot\int_{-\infty}^{+\infty}f(\delta\theta)\cos(\delta\theta)d\delta\theta}{p_++p_-+p_{11}}\\
&=V_{\theta}\cdot C{ph},
\end{aligned}
\end{equation}
where $C_{ph}=\int_{-\infty}^{+\infty}f(\delta\theta)cos(\delta\theta)d\delta\theta$ is a coefficient introduced by phase fluctuation, in which $f(\delta\theta)$ is the Gaussian probability density function of phase distribution. Through calculation, we get $C_{ph}=0.989$ and $0.973$ for 10~km and 50~km situation.

\subsubsection{Mismatch of write-out fields}
\begin{figure}[htb]
	\centering
	\includegraphics[width=0.5\textwidth]{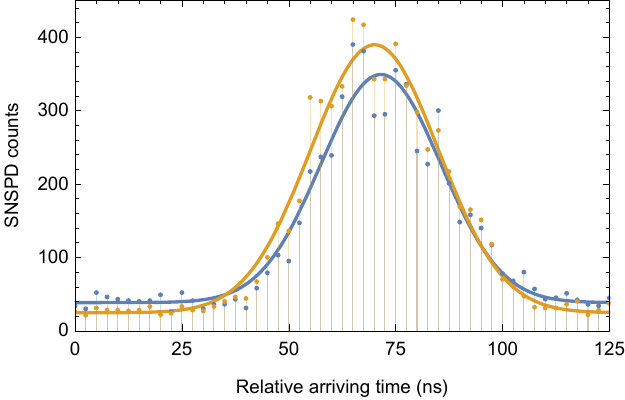}
	\caption{Time domain waveform of write-out field from two different MOTs. Dots for counts in SNSPD and curves for their fittings.}
	\label{fig:waveform}
	
\end{figure}

In entanglement building process, the write-out fields need to be calibrated to have the same arriving time. We adjust the difference of two optical paths length to achieve it. Via accumulating counts in SNSPD, we construct the shape of write-out fields and compare them. As depicted in Fig.~\ref{fig:waveform}, there exist a mismatch of 2.10~ns and 1.45~ns for 10~km and 50~km situation, which will bring $5.8\times 10^{-3}$ and $3.0 \times 10^{-3}$ decrease to $V_{\theta}$ separately.

\subsubsection{Decrease of SNR}

Along with write-out field being attenuated in long distance situation, noise of phase stabilization and dark counts of SNSPD introduce more disturbance. In 10~km and 50~km situation, we get about $15:1$ and $4.5:1$ signal to noise (SNR) ratio in SNSPD between with and without write-out field input.

First, in consideration of the definition of $V_{\theta}\equiv(max-min)/(max+min)$, random noise will contribute to $max$ and $min$ term equally. Therefore, we have disturbed visibility
\begin{equation}
\begin{aligned}
	V_{\theta}'=& \frac{p_{max}+p_{noise}-p_{min}-p_{noise}}{p_{max}+p_{noise}+p_{min}+p_{noise}}\\
	=& V_{\theta}\cdot\frac{c_{v}}{c_{v}+c_{n}}.
\end{aligned}
\end{equation}
$c$ refers to the coincidence probability with subscript denoting its source, $v$ to the valid combination, and $n$ to unwanted ones from noise. They could be estimated as:
\begin{equation}
  c_{v}\approx 2p_{wo}\eta_{ret}\ \  ,\ \   c_{n}\approx2p_n\cdot2p_r.
\end{equation}
$p_{wo}$ is the tested probability of write-out photons in SNSPSD. $p_n=p_{wo}/SNR$ is the noise in SNSPD. $p_r$ is the probability of read-out photons. After calculations, we know that $V_{\theta}$ suffers $0.003$ and $0.01$ decrease for 10~km and 50~km situation.

Second, the extra noises will contribute more vacuum parts in the remote entangled state. Considering the non-zero term in concurrence estimation
\begin{equation}\label{eq:concp}
\mathcal{C}^+=\frac{V_{\theta}(p_{01}+p_{10})-2\sqrt{p_{00}p_{11}}}{p_{00}+p_{01}+p_{10}+p_{11}}.
\end{equation}
The extra vacuum part will decrease the numerator and increase the nominator simultaneously. $p_{00}$ could be estimated as
\begin{equation}
	p_{00}\simeq (p_{01}+p_{10})\cdot\frac{1}{SNR}.
\end{equation}
Feeding this estimation into Eq.~\ref{eq:concp} we can calculate $\mathcal{C}^+$ as 0.57 and 0.45 for 10~km and 50~km situation. This result implies that the extra noise is the main reason of the decrease of $\mathcal{C}$ in long fiber situation.

\section{Entanglement evaluation of Fock state entanglement}
\subsection{Concurrence}
We consider the density matrix of two atomic ensembles or corresponding retrieved read-out fields as
\begin{equation}
\rho=\frac{1}{P}{
	\left( \begin{array}{cccc}
	p_{00}	 & 0	 & 0		& 0\\
	0   	 & p_{01}& d		& 0\\
	0 		 & d^*	 & p_{10}	& 0\\
	0		 & 0 	 & 0		& p_{11}\\
	\end{array}
	\right ).}
\end{equation}
It is in basis $\ket{n}_A\ket{m}_B$, where $\{n,m\}=\{0,1\}$ is the population numbers of spin-wave or photon. $p_{ij}$ is the corresponding probability and d gives the relative coherence. Although it is not a complete description of the whole Hilbert space, concurrence as an entanglement measure, its monotonicity under local operation and classical communication (LOCC)\cite{guhne2009-S}, which all imperfections in our experiment are involved, ensures that we always get the lower bound. (See Supplementary Information of ref.~\cite{Chou2005-S} for details.)

However, to approach the real value of concurrence as much as possible, we consider here the deduction of the impact from those clearly known influences. In order to detect spin-waves in atomic ensembles, we retrieve it into read-out photon. This process suffers from limited retrieval efficiency, loss in optical components and limited detection efficiency of APDs. One can model the loss, wherever it comes from, as an operator performed on every population of spin-wave or photon mode simultaneously:
\begin{equation}
   \openone\rightarrow\sqrt{\eta}\openone+\sqrt{1-\eta}a.
\end{equation}
$\openone$ and $a$ are identity and annihilation operator, respectively. Thus, we can build a relation between the detection probability $d_{ij}$ with the $p_{ij}$ in density matrix as:
\begin{equation}
	\left( \begin{array}{c}
	d_{00}	\\
	d_{01} 	\\
	d_{10} 	\\
	d_{11}	\\
	\end{array}
	\right )
	=
	{\left( \begin{array}{cccc}
	1		 & 1-\eta	 & 1-\eta	& (1-\eta)^2\\
	0   	 & \eta	  	&  0		& \eta(1-\eta)\\
	0 		 & 0 		 & \eta		& \eta(1-\eta)\\
	0		 & 0 		 & 0		& \eta^2\\
	\end{array}
	\right )}
	\left( \begin{array}{c}
	p_{00}	\\
	p_{01} 	\\
	p_{10} 	\\
	p_{11}	\\
	\end{array}
	\right ).
\end{equation}
By solving the equation we can trace the state back and calculate concurrence. Tab.~\ref{tab:con} shows the concurrence $\mathcal{C}_{raw}$ calculated by raw data, $\mathcal{C}_{ro}$ by deducing optical loss, which characterizes the read-out field, and $\mathcal{C}_{a}$ by deducing retrieval loss, which characterizes the atomic ensemble. Raw data of $d_{ij}$ is listed in Tab.~\ref{tab:res}.
\begin{table*}[htbp]
	\caption{Concurrence and estimated Fidelity in different situations in SPI experiment. $\bm{\mathcal{C}_{raw}}$, $\bm{\mathcal{C}_{p}}$ and $\bm{\mathcal{C}_{a}}$ refer to concurrence calculated by raw data, subtracting detection and optical loss and subtracting retrieval loss, respectively. $\bm{\mathcal{F}_{est}}$: estimated Fidelity of PME state.}
		\scriptsize
		\label{tab:con}
		\renewcommand\arraystretch{1.2}
		\begin{ruledtabular}
			\begin{tabular}{C{2.3cm} C{0.5cm} C{3cm} C{3cm} C{3cm} C{1cm}}
				& &  $\bm{\mathcal{C}_{raw}}$  &  $\bm{\mathcal{C}_{p}}$  &  $\bm{\mathcal{C}_{a}}$  & $\bm{\mathcal{F}_{est}}$ \\
				\hline
				\multirow{2}*{\textbf{Not conv./10~m}}& $\bm{D_a}$ & $0.151\pm 0.003$ & $0.484\pm0.008$ & $0.678\pm0.012$ & $0.786$ \\
				&$\bm{D_b}$ & $0.147\pm 0.003$ & $0.486\pm0.007$ & $0.711\pm0.012$ & $0.774$ \\
				\specialrule{0em}{3pt}{3pt}
				\multirow{2}*{\textbf{Conv./10~m}}&$\bm{D_a}$ & $0.155\pm 0.006$ & $0.484\pm0.015$ & $0.665\pm0.020$ & $0.781$ \\
				&$\bm{D_b}$ & $0.144\pm 0.006$ & $0.438\pm0.016$ & $0.581\pm0.020$ & $0.788$ \\
				\specialrule{0em}{3pt}{3pt}
				\multirow{2}*{\textbf{Conv./10~km}}&$\bm{D_a}$ & $0.092\pm 0.004$ & $0.311\pm0.009$ & $0.428\pm0.013$ & $0.701$ \\
				&$\bm{D_b}$ & $0.090\pm 0.001$ & $0.303\pm0.003$ & $0.416\pm0.008$ & $0.704$ \\
				\specialrule{0em}{3pt}{3pt}
				\multirow{2}*{\textbf{Conv./50~km}}&$\bm{D_a}$ & $0.088\pm 0.002$ & $0.297\pm0.005$ & $0.407\pm0.008$ & $0.692$ \\
				&$\bm{D_b}$ & $0.077\pm 0.003$ & $0.257\pm0.008$ & $0.348\pm0.011$ & $0.690$ \\
			\end{tabular}
	\end{ruledtabular}
\end{table*}

\begin{table*}[!htbp]
		\caption{Raw data of SPI experiment.}
		\scriptsize
		\label{tab:res}
		\renewcommand\arraystretch{1.2}
		\begin{ruledtabular}
			\begin{tabular}{C{2.3cm} C{0.5cm} C{2.5cm} C{2.5cm} C{2.5cm} C{2.5cm} C{2.5cm}}
				& &  $\bm{d_{00}}$  &  $\bm{d_{01}}$  &  $\bm{d_{10}}$  & $\bm{d_{11}}$ &$\bm{V_{\theta}}$ \\
				\hline
				\multirow{2}*{\textbf{Not conv./10~m}}& $\bm{D_a}$ & $0.6683\pm 0.0049$ & $0.1664\pm0.0032$ & $0.1599\pm0.0032$ & $0.0053\pm0.0005$&$0.8288\pm0.0085$ \\
				&$\bm{D_b}$ & $0.6588\pm 0.0052$ & $0.1728\pm0.0035$ & $0.1624\pm0.0034$ & $0.0061\pm0.0006$&$0.8168\pm0.0078$ \\
				\specialrule{0em}{3pt}{3pt}
				\multirow{2}*{\textbf{Conv./10~m}}&$\bm{D_a}$ & $0.6728\pm 0.0051$ & $0.1623\pm0.0034$ & $0.1607\pm0.0033$ & $0.0041\pm0.0005$&$0.8068\pm0.0187$ \\
				&$\bm{D_b}$ & $0.6963\pm 0.0049$ & $0.1560\pm0.0033$ & $0.1440\pm0.0032$ & $0.0037\pm0.0005$&$0.8194\pm0.0215$ \\
				\specialrule{0em}{3pt}{3pt}
				\multirow{2}*{\textbf{Conv./10~km}}&$\bm{D_a}$ & $0.7043\pm 0.0029$ & $0.1470\pm0.0019$ & $0.1444\pm0.0019$ & $0.0043\pm0.0003$&$0.6944\pm0.0128$ \\
				&$\bm{D_b}$ & $0.7084\pm 0.0032$ & $0.1463\pm0.0021$ & $0.1409\pm0.0021$ & $0.0044\pm0.0004$&$0.7037\pm0.0040$ \\
				\specialrule{0em}{3pt}{3pt}
				\multirow{2}*{\textbf{Conv./50~km}}&$\bm{D_a}$ & $0.7089\pm 0.0039$ & $0.1480\pm0.0026$ & $0.1391\pm0.0026$ & $0.0040\pm0.0004$&$0.6767\pm0.0065$ \\
				&$\bm{D_b}$ & $0.7287\pm 0.0042$ & $0.1394\pm0.0028$ & $0.1282\pm0.0027$ & $0.0038\pm0.0004$&$0.6787\pm0.0125$ \\
			\end{tabular}
	\end{ruledtabular}
\end{table*}

\subsection{Fidelity estimation}
\begin{figure}[!b]
	\centering
	
	\includegraphics[width=4.8in]{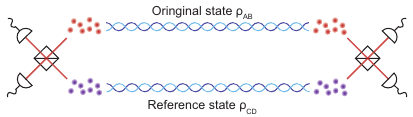}
	
	\caption{Illustration of next step in DLCZ protocol.}
	\label{fig:nextstep}
	
\end{figure}

In the next step of DLCZ protocol, another EME state $\rho_{CD}$ is introduced and leads to a PME state as shown in Fig.~\ref{fig:nextstep}. So it will be helpful to estimate the Fidelity of PME state with current data.
For simplicity of analysis, we rewrite the density matrix of two ensembles in another complete set of bases \{$\ket{00}$, $\ket{\Psi^+}$, $\ket{\Psi^-}$, $\ket{11}$\} as
\begin{equation}\label{eq:mix}
\rho_{AB}^{ro}=\frac{1}{P}(p_{00}\ket{00}\bra{00}+p_{+}\ket{\Psi_p^+} \bra{\Psi_p^+}+p_-\ket{\Psi_p^-}\bra{ \Psi_p^-}+p_{11}\ket{11}\bra{11}+{A.D.}),
\end{equation}
where $\ket{\Psi^{\pm}}=(\ket{01}\pm\ket{10})/\sqrt{2}$ and $A.D.$ refers to anti-diagonal terms. It is easy to get $p_{\pm}=(1\pm V)(p_{01}+p_{10})/2$.

Four ensembles are retrieved into read-out fields simultaneously and two fields from one side are combined together by a BS. Through registering coincidence that only one detector clicks on each side, we capture PME state and perform communication. We consider this process and suppose an ideal state $\rho_{CD}=|\Psi^+\rangle\langle\Psi^+|$ first for simplicity. Here we list four items of $\rho_{AB}\otimes\rho_{CD}$ and their contributions in next step in Tab.~\ref{tab:nextstep1}.

\begin{table*}[htbp]
	\caption{\label{tab:nextstep1}All combinations in PME preparation process with $\rho_{AB}$ and $|\Psi^+\rangle\langle\Psi^+|$.}
	\scriptsize
	\renewcommand\arraystretch{1.5}
	\begin{ruledtabular}
	\begin{tabular}
		{cccccc}
		&
		&$|00\rangle$&$|\Psi^-\rangle$&$|\Psi^+\rangle$&$|11\rangle$ \\
		\hline
		\multicolumn{2}{c}{\textbf{Probability}}
		&  $p_{00}$  &  $p_{+}$  &  $p_{-}$  &  $p_{11}$ \\
		\multirow{2}*{\textbf{Coef.}} &\textbf{Right}
		&  0  &  1/2  &  0  &  1/2 \\
		&\textbf{Wrong}
		&  0  &  0  &  1/2  &  1/2 \\
	\end{tabular}
\end{ruledtabular}
\end{table*}

The vacuum part contributes nothing to final results due to no coincidence it gives. Coefficient $1/2$ from $|\Psi^+\rangle$ is a intrinsic success probability of this protocol. Because of the converse phase, $|\Psi^-\rangle$ gives rise to false coincidences. Owing to two read-out fields existing at the same time, $|11\rangle$ always gives coincidences. And it contribute equally to right and wrong part thanks to randomness. We calculate the fidelity of PME state as
\begin{equation}
\mathcal{F}_{\rho,+}=\frac{Right}{Right+Wrong}=\frac{p_++p_{11}}{p_++p_{-}+2p_{11}}
\end{equation}
Considering more realistic situation that $\rho_{CD}=\rho_{AB}$. We list all outputs in Tab.~\ref{tab:nextstep2}. Then the fidelity of PME state could be calculated similar as above. We show the simulation results in Tab.~\ref{tab:con}. Besides, the estimated Fidelity $\mathcal{F}_{est}$ hardly varies whether we subtract the loss or not, implying a good filtering result to the vacuum part of the DLCZ protocol.

\begin{table*}[htbp]
\caption{\label{tab:nextstep2}All combinations in PME preparation process with two $\rho_{AB}$s.}
	\scriptsize
	\renewcommand\arraystretch{1.5}
	\begin{ruledtabular}
	\begin{tabular}
		{cccccccccccccccccc}
		&
		&$|00\rangle$&$|00\rangle$&$|00\rangle$&$|00\rangle$ &$|\Psi^+\rangle$&$|\Psi^+\rangle$&$|\Psi^+\rangle$&$|\Psi^+\rangle$
		&$|\Psi^-\rangle$&$|\Psi^-\rangle$&$|\Psi^-\rangle$&$|\Psi^-\rangle$
		&$|11\rangle$&$|11\rangle$&$|11\rangle$&$|11\rangle$\\
		 &
		 &$|00\rangle$&$|\Psi^+\rangle$&$|\Psi^-\rangle$&$|11\rangle$ &$|00\rangle$&$|\Psi^+\rangle$&$|\Psi^-\rangle$&$|11\rangle$
		 &$|00\rangle$&$|\Psi^+\rangle$&$|\Psi^-\rangle$&$|11\rangle$
		 &$|00\rangle$&$|\Psi^+\rangle$&$|\Psi^-\rangle$&$|11\rangle$\\
		\hline
		 \multicolumn{2}{c}{\textbf{Probability}}
		 &$p_{00} p_{00}$&$p_{00} p_{+}$&$p_{00} p_{-}$&$p_{00} p_{11}$
		 &$p_{+} p_{00}$ &$p_{+} p_{+}$ &$p_{+} p_{-}$ &$p_{+} p_{11}$
		 &$p_{-} p_{00}$ &$p_{-} p_{+}$ &$p_{-} p_{-}$ &$p_{-} p_{11}$
		 &$p_{11} p_{00}$&$p_{11} p_{+}$&$p_{11} p_{-}$&$p_{11} p_{11}$
		  \\
		\multirow{2}*{\textbf{Coef.}} &\textbf{Right}
		&  0  &  0    &  0  &  1/2
		&  0  &  1/2  &  0  &  1/2
		&  0  &  0    & 1/2 &  1/2
		& 1/2 &  1/2  & 1/2 &  1/2
		 \\
		&\textbf{Wrong}
		&  0  &  0    &  0  &  1/2
		&  0  &  0    & 1/2 &  1/2
		&  0  &  1/2  &  0  &  1/2
        & 1/2 &  1/2  & 1/2 &  1/2
		  \\
	\end{tabular}
\end{ruledtabular}
\end{table*}

\end{document}